\documentclass[preprint]{aastex63}

\shorttitle{Low-Redshift FeLoBALQs}
\shortauthors{Leighly et al.}

\begin{document}

\title{The Physical Properties of Low-Redshift FeLoBAL Quasars: II.\ The
  Rest-Frame   Optical Emission Line Properties}

\author[0000-0002-3809-0051]{Karen M.\ Leighly}
\affiliation{Homer L.\ Dodge Department of Physics and Astronomy, The
  University of Oklahoma, 440 W.\ Brooks St., Norman, OK 73019, USA}

\author[0000-0002-3173-1098]{Hyunseop Choi}
\affiliation{Homer L.\ Dodge Department of Physics and Astronomy, The
  University of Oklahoma, 440 W.\ Brooks St., Norman, OK 73019, USA}

\author{Cora DeFrancesco}
\affiliation{Homer L.\ Dodge Department of Physics and Astronomy, The
  University of Oklahoma, 440 W.\ Brooks St., Norman, OK 73019, USA}

\author{Julianna Voelker}
\affiliation{Homer L.\ Dodge Department of Physics and Astronomy, The
  University of Oklahoma, 440 W.\ Brooks St., Norman, OK 73019, USA}

\author{Donald M.\ Terndrup}
\affiliation{Department of Astronomy, The Ohio State University, 140
  W. 18th Ave., Columbus, OH 43210}
\affiliation{Homer L.\ Dodge Department of Physics and Astronomy,
  The
  University of Oklahoma, 440 W.\ Brooks St., Norman, OK 73019, USA}

\author{Sarah C.\ Gallagher}
\affiliation{Department of Physics \& Astronomy, The University of
  Western Ontario, London, ON, N6A 3K7, Canada}
\affiliation{Canadian Space Agency, 6767 Route de l'A\'eroport,
  Saint-Hubert, Quebec, J3Y BY9}
\affiliation{Institute for Earth and Space Exploration, The
  University of Western Ontario, London, ON, N6A 3K7, Canada}
\affiliation{The Rotman Institute of Philosophy, The University of
  Western Ontario, London, ON, N6A 3K7, Canada}

\author{Gordon T.\ Richards}
\affiliation{Department of Physics, Drexel University, 32 S. 32nd St.,
  Philadelphia, PA 19104}



\begin{abstract}
We report the results of analysis of the H$\beta$ emission-line region
of a sample of thirty low-redshift ($z<1$) 
iron low-ionization broad absorption line quasars (FeLoBALQs).
{ Eleven} of these objects are newly classified as FeLoBALQs.  A
matched
sample of 132 unabsorbed quasars was analyzed in parallel.   The
emission lines showed the well known anticorrelation between the
[\ion{O}{3}] and \ion{Fe}{2} emission \citep{bg92}.   Using a summary
statistic called $E1$ to quantify this anticorrelation, we found that
while the distribution of $E1$ for the unabsorbed quasars has a single
peak, the FeLoBALQs have a bimodal shape in this parameter.
Previous studies have shown that the line emission properties of BAL
and non-BAL quasars  
are consistent, and therefore the difference in the H$\beta$ region
emission between FeLoBAL quasars and unabsorbed quasars is a new
result. The two populations of FeLoBAL quasars are characterized by
low and high bolometric luminosities and Eddington ratios.  Some  
previous studies have suggested 
that BAL quasars are high accretion-rate objects, and therefore the
discovery of the low accretion-rate branch of FeLoBAL quasars was
unexpected. We also found that the H$\beta$ FWHM is systematically
broader among the FeLoBALQs compared with the non-BAL quasars implying
a higher inclination viewing angle or a dearth of low-velocity
line-emitting gas.

\end{abstract}



\section{Introduction} \label{intro}

Broad absorption lines occur in 10-26\% of optically selected
quasars \citep{tolea02, hf03, reichard03, trump06, knigge08,
  gibson09}.  The broad
and 
blueshifted \ion{C}{4} absorption lines observed in 
broad absorption line quasars (BALQs) reveal
an unambiguous signature of outflow.  Therefore, BALQs may be
important sources of quasar feedback in galaxy evolution.
\citet{trump06} found that 1.3\%
of quasars have broad \ion{Mg}{2} absorption; these
are called low-ionization broad absorption line quasars
(LoBALQs). {  A typical median width of a LoBALQ   \ion{Mg}{2} broad
absorption line is 4400 km s$^{-1}$ \citep{yi19}.}  
 \citet{trump06} found that 0.3\% of quasars also have
absorption from \ion{Fe}{2}, and these are called iron low-ionization
broad absorption line quasars (FeLoBALQs).  FeLoBAL
quasars are rare, 
and previous to the Sloan Digital Sky Survey, only a few  were known.
For example, \citet{hall02} 
published spectra and discussed the wide range of features observed in
23 unusual objects discovered in the SDSS;  many of those
objects are FeLoBAL quasars.  

How do BAL quasars, and FeLoBAL quasars specifically, fit in among
quasars in general?  Are they fundamentally the same, and their
magnificent spectra are observed because of a select range of viewing
angles?  Or do they mark a special stage in quasar evolution?  Or are
both factors important? 

One way to address these questions is to look at the broad emission
lines. \citet{weymann91} performed the first comprehensive study of the
emission-line properties of BAL and unabsorbed quasars, focusing on
the rest frame  UV emission lines observed in ground-based spectra.  They
found that, with the exception of LoBAL quasars, the line emission was 
indistinguishable from that of unabsorbed quasars.  This result makes
sense if the emission lines in both type of objects originate in a
similarly photoionized broad-line region (rather than, e.g., an
expanding outflow like a supernova).  Modern studies using larger
samples have found a somewhat different pattern of behaviors that link
the outflow properties with the emission line and continuum
properties.  For example, \citet{baskin15} reported that the
\ion{He}{2}$\lambda 1640$ equivalent width is 
inversely related to the velocity shift and width of the
BAL absorption. This result has been confirmed by further analysis
\citep{hamann19, rankin20}.  Because weak \ion{He}{2} is a signature
of a UV-dominant or soft spectral energy distribution \citep[SED;
  e.g.,][]{leighly04}, these results  
may imply that the illuminating SED in at least some BAL quasars tends
to be soft.   

The rest-frame optical bandpass is arguably the best-studied
region of quasar spectra, containing the H$\beta$, [\ion{O}{3}], and
\ion{Fe}{2} lines that are routinely used to measure black hole masses
and Eddington ratios.  The rest-frame optical band has been less well
studied in BAL quasars  than the rest-frame UV band, because this band
is observed in the near-infrared when \ion{C}{4}$\lambda 1549$ is
observed in the  optical band.  \citet{yw03} 
reported results from 16 BAL quasars and a comparison sample of 13
non-BAL quasars.  They found  that the [\ion{O}{3}] lines are weak in
BAL quasars.  \citet{runnoe13} reported results from a sample of eight
moderate-redshift radio-loud BALQs.  They also found weak [\ion{O}{3}]
and strong \ion{Fe}{2} in these objects suggesting a high accretion
rate \citep[e.g.,][]{boroson02, shen_ho_14}.  However, analysis of the
Balmer-line widths and estimation of the black hole masses revealed
the Eddington ratios that were consistent with comparable unabsorbed
quasars.   
\citet{schulze17} presented the largest sample of LoBAL and FeLoBAL
quasars studied in the rest-frame optical band, with 
 near-IR spectroscopy from 16 LoBAL quasars and 6 FeLoBAL
quasars.  Sixteen of the spectra cover the H$\beta$/[\ion{O}{3}]
region.  Their results deviate from the previous  work:  they saw
neither notably weak [\ion{O}{3}] nor notably strong \ion{Fe}{2}
emission.  Instead, they found that their $z \sim 1.5$ subsample
composite spectrum may have weaker \ion{Fe}{2} emission than the
unabsorbed quasars. 

These three studies of the rest-frame optical spectra of
BAL quasars leave the impression that BAL quasars often appear to be
high Eddington ratio objects.  More recently, \citet{rankin20} estimated
the Eddington ratio in BAL quasars using a reconstructed \ion{C}{4}
emission line and a profile correction \citep{coatman17}.  They found 
that the Eddington ratio distribution is indistinguishable between
broad absorption line quasars and unabsorbed quasars.  

A complicating factor is the luminosity of the objects studied.
Many quasar properties are luminosity dependent \citep[e.g., the
  Baldwin effect;][]{baldwin77}.   BAL outflow velocities are
observed to have a particularly strong dependence on quasar luminosity
\citep[e.g.,][]{laor02,ganguly07}.    This dependence
may be expected; for a fixed $L_\mathrm{Bol}/L_\mathrm{Edd}$, a more luminous black hole
will have a larger black hole mass.  That larger black hole mass
yields a spectral energy distribution (SED) that peaks at longer
wavelengths. This softer SED may influence the outflows in two ways.
1.) Soft SEDs are less likely to over-ionize the outflowing gas.  2.)
The soft SED produces a relatively higher flux density of the UV
photons responsible for accelerating the outflow via resonance line
driving.  Quasar luminosity evolution and flux-limited surveys combine
to yield higher-luminosity objects at larger redshifts
\citep{jester05}. It is therefore not clear that samples of objects
with different redshifts and luminosities can be compared directly.  

This paper is the second in a series of four papers.  Paper I
\citep{choi22} describes the {\it SimBAL} \citep{leighly18} spectral 
synthesis analysis of the BAL outflows in fifty low redshift FeLoBAL
quasars.  In this paper, Paper II, we present analysis of the
rest-frame optical band spectra of a subsample of thirty of the 50
FeLoBAL quasars considered in \citet{choi22}.  All of the objects in
this subsample have redshifts $z<1$ and the SDSS/BOSS spectra include
the H$\beta$ / [\ion{O}{3}] region of the spectrum.  Paper III (Choi et
al.\ submitted)  combines the {\it   SimBAL}  results from Paper I and the
emission-line analysis of this paper to discuss the
implications for the location and geometry of the outflow.  Paper IV
(Leighly et al.\ in prep.) includes the broad-band optical/IR properties and
discusses the potential implications for accretion models and
evolution scenarios for low redshift FeLoBAL quasars.

Our paper is organized as follows.  In \S\ref{data}, we describe the
data selection of the FeLoBAL quasars and a matched comparison sample
of 132 unabsorbed quasars, the optical band modeling, and a principal
components analysis of the spectrum around H$\beta$.  In
\S\ref{dist_comp}, we compare the optical and derived properties of
the FeLoBAL quasars with those of the comparison sample.  We present
our reasoning and methodology for dividing the FeLoBAL quasars into
two groups.  We also look at correlations and patterns among the
emission-line and global properties and compare with
previous results from the literature.  \S\ref{discussion} summarizes
our results.

\section{Data} \label{data}

\subsection{Sample Selection} \label{selection}

\begin{figure*}[!t]
\epsscale{1.0}
\begin{center}
\includegraphics[width=4.5truein]{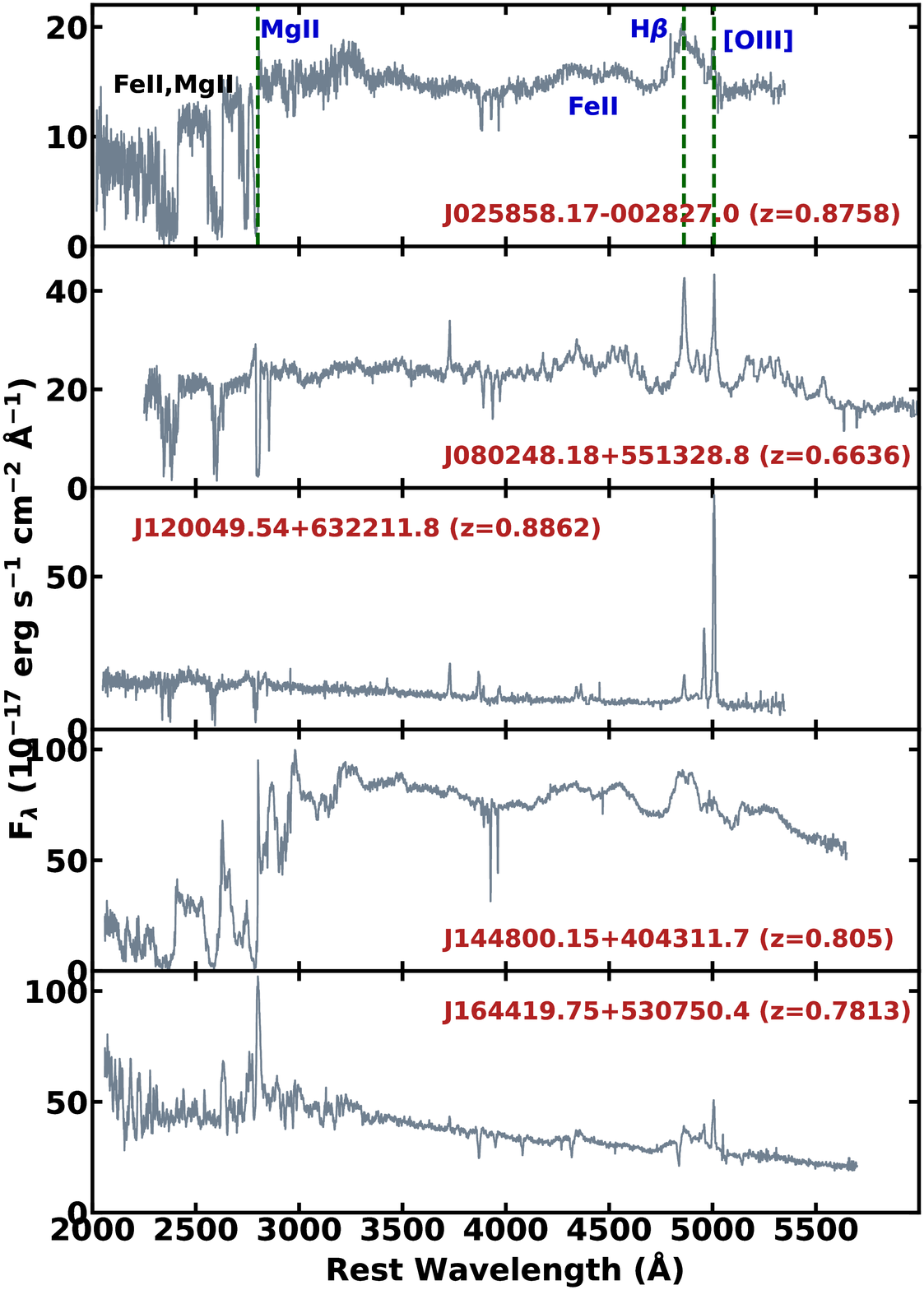}
\caption{The SDSS or BOSS spectra of five FeLoBALQs from our sample.
  The   \ion{Fe}{2} and \ion{Mg}{2} absorption can be seen at short
  wavelengths, and the H$\beta$, [\ion{O}{3}], and \ion{Fe}{2} complex
  is present at long wavelengths.  { The vertical dashed lines in the
  top line show the rest-frame wavelengths of \ion{Mg}{2}, H$\beta$,
  and the 5007\AA\/ component of [\ion{O}{3}].}  A wide range of absorption and
  emission morphologies can be seen. The optical-band spectra of the
  full sample are shown in Fig.~\ref{fig1a}, while the near-UV spectra 
  (with {\it SimBAL} model fits) are found in \citet{choi22}.
   \label{example_spectra}}
\end{center}
\end{figure*}

The parent sample of the FeLoBALQs was drawn from the literature, from
the sample of $0.8 < z < 3.0$ FeLoBALQ candidates found in the DR14
quasar catalog \citep{paris18} using a convolutional neural net
trained with synthetic spectra \citep[][Dabbieri et al.\ in
  prep.]{dabbieri20}, and from visual inspection.  The redshift range of
the objects was principally between $\sim 0.7$ and 1.0; several of the
objects with redshifts smaller than $z=0.75$ are  well-known
FeLoBALQs \citep[FBQS~J104459.5$+$365605,
  FBQS~J1214$+$2803,][]{dekool01,dekool02b}.  The resulting 
sample included thirty objects that had spectra of sufficient quality
to analyze both the BAL absorption at the short wavelength end and
the H$\beta$ and [\ion{O}{3}] emission at the long wavelength end.
{ The other twenty objects analyzed by \citet{choi22}
  had too high redshift to include the rest-frame H$\beta$ and
  [\ion{O}{3}] emission at the long-wavelength end.}
The spectra of five example FeLoBALQs from the sample displayed in 
Fig.~\ref{example_spectra} illustrate the wide range of absorption
line and emission line properties in the sample.  The properties of
the sample are presented in Table~1 of \citet{choi22}.

{ The sample of thirty objects can be divided into three classes
depending the origin of their BAL classification.  Ten objects are
classified as BAL quasars in the NASA/IPAC Extragalactic
Database\footnote{NED: http://ned.ipac.caltech.edu}.  Nine more are
included in the SDSS DR12 BAL quasar catalog \citep{paris17}.  The
remaining eleven are newly classified as BAL quasars. }

\begin{figure*}[!t]
\epsscale{1.0}
\begin{center}
\includegraphics[width=6.5truein]{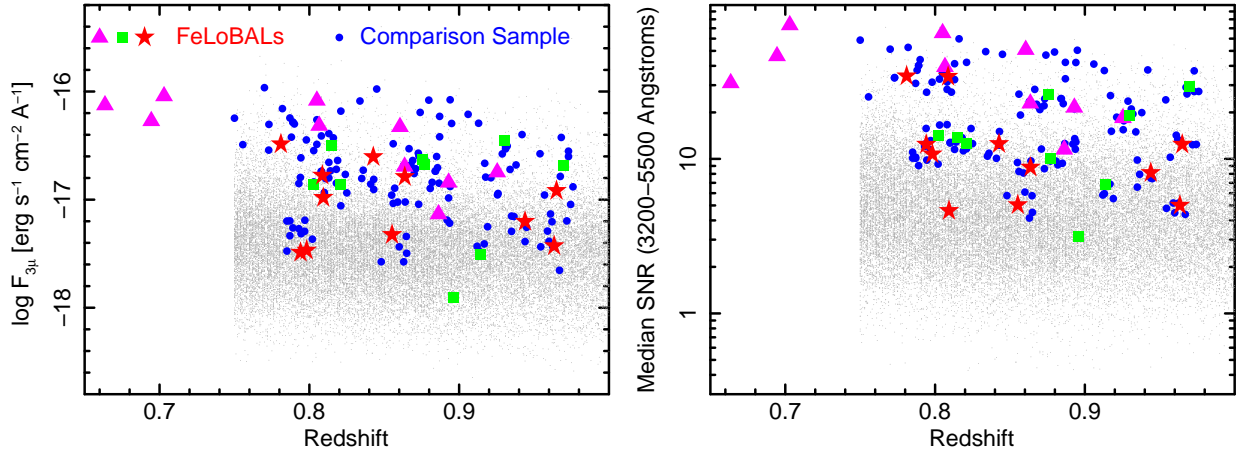}
\caption{Distribution of the FeLoBAL quasar and comparison sample as a
  function of the comparison-sample matching properties: the 3$\mu m$
  flux density, the redshift, and the median SNR in the
  rest frame 3200--5500\AA\/ bandpass. { The FeLoBALQs are subdivided
  into objects classified as BAL quasars in NED (magenta triangles),
  objects included in the SDSS DR12 BAL catalog \citep[green
    squares][]{paris17}, and objects newly classified as BAL quasars
  (red stars). }  The redshift-selected parent sample,
  shown in grey, was  drawn from the DR14 quasar catalog
  \citep{paris18}.  The three  objects at low redshift are
  previously identified, well-known   FeLoBAL quasars.
   \label{sample_dist}} 
\end{center}
\end{figure*}

{ For comparison,} we created a sample of unabsorbed
quasars.  Our goal was to identify five unabsorbed quasars from the
SDSS DR14Q \citep{paris18} for each FeLoBAL quasar.  For each
FeLoBALQ, we selected a set of unabsorbed objects that lay near it in
redshift, 3-micron luminosity density, and signal-to-noise ratio. {
 Our criteria for proximity were $\Delta \log F_\mathrm{3\mu}=0.3$, $\Delta
 z=0.01$, and $\Delta SNR=0.1$.}
{ Many quasar properties depend on luminosity; examples include the
  emission-line equivalent widths \citep{baldwin77} and BAL velocity
  \citep[e.g.,][]{ganguly07}.  BAL quasars tend to be reddened
  \citep[e.g.,][]{krawczyk15}, and evidence for reddening is present
  in this sample (Leighly et al.\ in prep.).  Therefore, we use the
  3-micron luminosity as representative, rather than the luminosity in
  the optical or UV. }
From those, we randomly chose 5 objects, visually inspecting each one
to be sure that it was not a BALQ and had analyzable
H$\beta$/[\ion{O}{3}] emission lines, { i.e., the H$\beta$ line
  could be discerned by eye.}  Six objects had such high
signal-to-noise ratios that this selection method yielded too few
comparison objects.  For those, the redshift criterion was
relaxed. Finally, the three objects with the poorest signal-to-noise
ratio did not yield sufficient analyzable comparison sample;
regardless, we retained these objects in our FeLoBALQ sample.  The
result was a comparison sample of 132 unabsorbed quasars.   The
distribution of the sample properties is shown in
Fig.~\ref{sample_dist}.    

\subsection{Optical Band Modeling} \label{optical_modeling}

Our goals for the optical-band modeling were to fine tune the
redshifts (necessary to measure the outflow velocities), 
extract the H$\beta$  properties to estimate the black hole mass, and
extract the [\ion{O}{3}] and \ion{Fe}{2} properties.  We modeled the
optical-band spectra using {\tt Sherpa}\footnote
{https://sherpa.readthedocs.io/en/latest/} \citep{freeman01}. 
  Although intended for spectral fitting in the X-ray band,
  {\tt Sherpa} has more than sufficient flexibility and robustness to 
  fit optical spectra.  

In order to take into account potential covariance between emission
lines, continuum, and the \ion{Fe}{2} pseudo-continuum, we modeled all
components simultaneously.    Most of the spectra were modeled between
3550\AA\/ and 5500\AA\/. Some Seyfert 1.5 and 1.8 objects were modeled
down to 3300\AA\/ to include the [\ion{Ne}{5}] lines. Sometimes the 
long-wavelength limit was as low as 5200\AA\/ for higher-redshift
objects due to the long wavelength limit of the SDSS/BOSS
spectrograph.  

Estimating the redshift from optical emission lines is known to be
difficult because many lines show evidence for outflows.   Following
the discussion in \citet{zg14}, we usually used the lowest ionization
narrow line available to estimate  the redshift.  When  H$\beta$
could be clearly distinguished from broad H$\beta$, we  used that.
{ Narrow H$\beta$ was fit in 16 out of 30 of the spectra.  Most
commonly, we used [\ion{O}{2}]$\lambda 3727$, which was present in 27
out of 30 of the spectra.}  When neither line was available, the SDSS 
DR14 redshift was used, or the redshift was estimated from the shape
of the near-UV \ion{Fe}{2} emission.

The continuum under the H$\beta$ and [\ion{O}{3}] lines was modeled
using a power law.  In some cases, we observed an upturn of the
spectrum towards shorter wavelengths.  We modeled that upturn with
either a recombination continuum model or a broken power law model.  

Most frequently, we modeled the \ion{Fe}{2} emission using the line
lists developed from the strong-\ion{Fe}{2} Seyfert galaxy I~Zw~1
\citep{vcjv04}.  
In a few cases, the
\ion{Fe}{2} emission is very prominent and complex, and we used the
more flexible \ion{Fe}{2} model constructed by \citet{kovacevic10}.
This model is comprised of five line lists, three of which correspond
to different lower levels of the transitions.  This model only spans 
4400--5500\AA\/, and we modeled the 3550--4400\AA\/ region using the
\citet{vcjv04} model, fitting both segments of the spectrum
simultaneously.  

The Balmer lines (H$\beta$, H$\gamma$, H$\delta$, H$\epsilon$) were
modeled using either a single Lorentzian profile, a single Gaussian
profile, or two Gaussians.  When a cusp could be seen in the H$\beta$
profile, the broad lines were modeled with Gaussians, and a set of narrow
Gaussian lines were included to model the narrow Balmer emission.
There should be a contribution to the H$\beta$ profile from the
narrow-line region, but we did not model it unless a cusp was present.
We experimented with including a narrow H$\beta$ component with width
and position tied to the [\ion{O}{3}] lines and an intensity one tenth
of the [\ion{O}{3}] emission \citep{cohen83}, and found that the
measurements of H$\beta$ agreed within the error bars with the results
of models that include a scaled narrow component. The
2-broad-Gaussian profile was only required for a few objects, and was 
most frequently used when the H$\beta$ line appeared to have a red
wing.  

We modeled the [\ion{O}{3}] lines with either one or two pairs of
Gaussian profiles.  As usual, the 4959\AA\/  component was
constrained to have the same width, 2.92 times the intensity, and lie a
fixed wavelength ratio from the 5007\AA\/ component.  In most
cases, the preferred model used two pairs of lines; however, one pair
was used when the line was particularly weak (especially when \ion{Fe}{2}
was strong).

Several objects (J1030$+$3120, J1044$+$3656, J1214$+$2803,
J1448$+$4043) were particularly challenging to model because of a
combination of good signal-to-noise ratio, strong \ion{Fe}{2}, and
weak [\ion{O}{3}].  One of the complicating factors is that we
measured statistically significant blueshifts in the  \ion{Fe}{2} emission
in J1214$+$2803 and J1448$+$4043 (by 330 and 740 $\rm km\, s^{-1}$,
respectively).  While the cosmological redshift of J1448$+$4043
remains uncertain because there are no low-ionization NLR reference
lines, [\ion{O}{2}] was observed in the J1214$+$2803 spectrum, and the
redshift 
could be measured securely.  To measure the emission-line properties
of these four objects, we assumed that they are physically similar,
and then used the values of the parameters that could be measured in
some objects to model other objects.  Specifically,  [\ion{O}{3}]
could be fully constrained by a single pair of gaussians in position,
intensity, and width in both J1030$+$3120 and J1044$+$3656.  The
[\ion{O}{3}] was too obscured to be fully constrained in J1214$+$2803
and J1448$+$4043.  The median value of the [\ion{O}{3}] FWHM for
J1030$+$3120 and  J1044$+$3656  was approximately $2000\rm \, km\,
s^{-1}$.  So for J1214$+$2803 and J1448$+$4043, the FWHM was fixed at
that value, and the intensity and position were modeled.  The plots of
the spectra and their model fits are given in Fig.~\ref{fig1a}.  The
measured values of emission-line parameters are given in
Table~\ref{data}.  

{ In some objects, the \ion{Fe}{2} and/or [\ion{O}{3}] emission was
weak.  We used the F test \citep{bevington} to gauge the statistical
signficance of these parameters in each model for both the FeLoBAL
quasar and comparison sample.  We used a significance cutoff of
$p=0.05$.  We found that 40 and 15 (30\% and 50\%) of the comparison and
FeLoBALQ sample did not statistically require the \ion{Fe}{2}
emission.  We found that 12 and 8 (16\% and 27\%) of the comparison and
FeLoBALQ sample did not statistically require the [\ion{O}{3}]
emission.} 

No stellar population was apparent in the spectra of most 
of the FeLoBAL and comparison sample objects, so we did not
include a young stellar population in our model.  We reasoned 
that because the signal-to-noise ratio of many of our spectra is low,
an additional  continuum component  could  not be robustly constrained.
Not including this component, assuming that it is present in some
objects, adds scatter to the equivalent widths and velocity width of
both the FeLoBALQ and comparison samples.   

Several emission-line properties were derived from the
optical-model-fitting results; these are compiled in Table~\ref{data}
for the FeLoBAL quasars.  
The total [\ion{O}{3}] luminosity was computed 
from the [\ion{O}{3}]$\lambda 5007$ line flux alone.  
We computed the parameter R$_\mathrm{FeII}$, commonly defined
as the ratio of the \ion{Fe}{2} equivalent width between 4434--4684
\AA\/ to the broad  H$\beta$ equivalent width.  We followed the
methodology of \citet{zg14} to extract the [\ion{O}{3}] median
velocity offset $v_{50}$ and velocity width $w_{80}$.  Briefly, from
the normalized cumulative function of the broad H$\beta$ model
profile, the velocities at 0.1, 0.5, and 0.9 were identified. The
velocity at 0.5 is assigned to $v_{50}$, and $w_{80}$ is the
difference between the velocities at 0.1 and 0.9.   It is
important to note that, unlike \citet{zg14} and similar studies of
obscured quasars, the quality of these measurements varied
dramatically through our sample, and are not very robustly measured in
objects with very strong \ion{Fe}{2} and weak [\ion{O}{3}].   

We estimated the bolometric luminosity using the estimate of the
rest-frame flux density at 3 microns and the bolometric correction
derived by \citet{gallagher07}. { This bolometric
  correction was derived from 259 quasars with $0.14 < z < 5.22$ and
  estimated log bolometric luminosities $45.1 < \log L_\mathrm{Bol} <
  47.4$ [erg s$^{-1}$] with a median value of 46.4.  Our estimated
  bolometric luminosities fall between $45.1 < 
  \log L_\mathrm{Bol} < 46.2$ with a median value of 46.2. Although
  our objects are somewhat less luminous than the quasars considered
  by \citet{gallagher07}, the  difference in median is only 0.2 dex.}
  We obtained the 3-micron flux density
from the WISE photometry by doing a log-linear interpolation.  For
objects in this redshift range (0.7--0.97), 3 microns falls near the
W2 effective wavelength (4.6 microns), and generally the W2 magnitude
was measured with good signal-to-noise ratio in this sample of
relatively nearby objects.  

The black hole mass estimate requires an estimate of the radius of the
H$\beta$-emitting broad-line region.  We used the formalism given by
\citet{bentz06}.  The flux density at 5100\AA\/ was calculated from
the power-law model component of the emission-line model fits, and
then used to estimate the radius of the H$\beta$ 
emitting region.  We did not correct for reddening
intrinsic to the quasar, and therefore objects with
significantly-reddened spectra will have somewhat 
underestimated black hole masses.  The black hole mass was estimated
using the FWHM of the H$\beta$ line and following the formalism of
\citet{collin06}.  In particular, we used their Eq.\ 7 to estimate the 
scale factor $f$ based on the FWHM of the H$\beta$ line.  

We calculated the size of the 2800\AA\/ continuum emission region
using the estimated black hole mass and the bolometric luminosity and 
the procedure described in \S~6.1 of \citet{leighly19}.  To
summarize, we used a simple sum-of-blackbodies accretion disk model
\citep{fkr02} and assigned the 2800\AA\/ radius to be the location
where the radially-weighted brightness dropped by a factor of $e$ from
the peak value.  This estimate may not be accurate for the full range
of Eddington ratios represented by the sample, since structure of the
accretion disk is thought to be different for very low and very high
accretion rates \citep[e.g.,][]{gp19}.

The uncertainties in the derived parameters were estimated using the
{\tt Sherpa} model fit errors.  Because the errors produced by  {\tt
  Sherpa} were sometimes asymmetric, we used a Monte 
Carlo scheme to propagate errors.   Specifically, we created 10,000
instances of each parameter varying the value within the errors, then
computed the derived property, and extracted the $1 \sigma$ errors from
the cumulative distributions.

{ We investigated whether there are any systematic differences in the
properties measured in this paper between the previously-classified
and the newly-classified BAL quasars using 
two-sample Kolmogorov-Smirnov  (KS) test and the two-sample
Anderson-Darling (AD) test.  Among those seventeen properties, we
found significant ($p<0.05$) differences in only three properties.  The
SPCA Eigenvector 4 coefficient (\S\ref{pca}) was lower in the new
objects.  Examining the plot of the PCA eigenvectors
(Fig.~\ref{pca_eigen}) shows that this means that objects with narrow
\ion{Fe}{2} emission lines are preferentially represented among the
previously-identified objects.  The median bolometric luminosity was
0.44 dex lower among the new FeLoBAL quasars compared with the
previously-known objects.  This reflects the fact that the brightest
objects in a sample are usually identified first, likely because they
have higher signal-to-noise ratio spectra.  Finally, the size
of the 2800\AA\/ emission region of the new objects 0.1 dex lower than
the previously-known objects, likely a consequence of the lower
luminosities.  }

The 132 comparison-sample spectra were subjected to the same analysis.  

\begin{figure*}[!t]
\epsscale{1.0}
\begin{center}
\includegraphics[width=6.0truein]{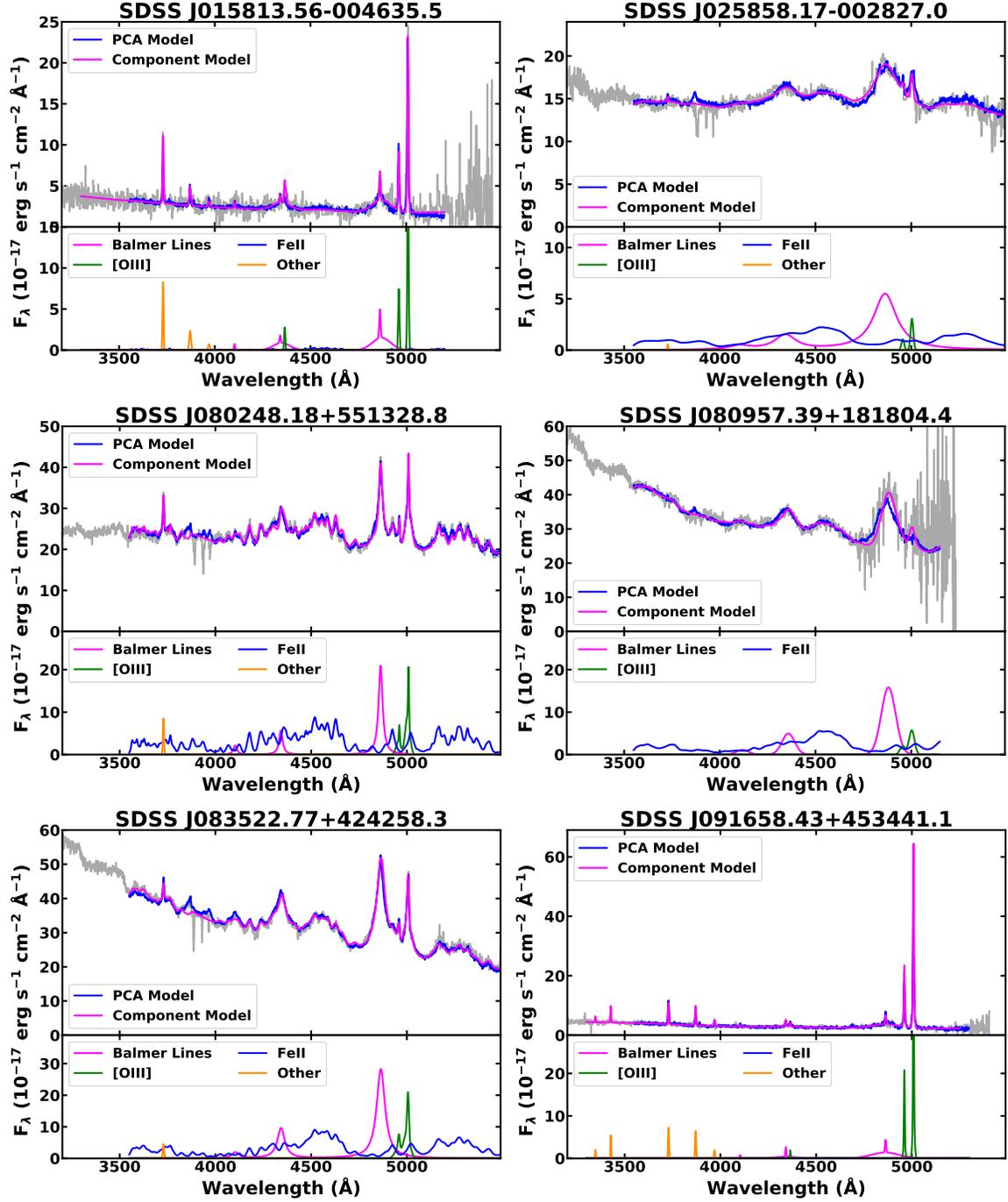}
\caption{The optical portion of the FeLoBAL sample spectra overlaid
  with the best-fitting model spectrum.  Both the component-based
  emission-line (\S\ref{optical_modeling}) and the principal component
  eigenvector  (\S\ref{pca}) model fits  are shown.  The lower panel
  shows the components from the   component-based model.  \label{fig1a}} 
\end{center}
\end{figure*}

\addtocounter{figure}{-1} 
\begin{figure*}[!t]
\epsscale{1.0}
\begin{center}
\includegraphics[width=6.0truein]{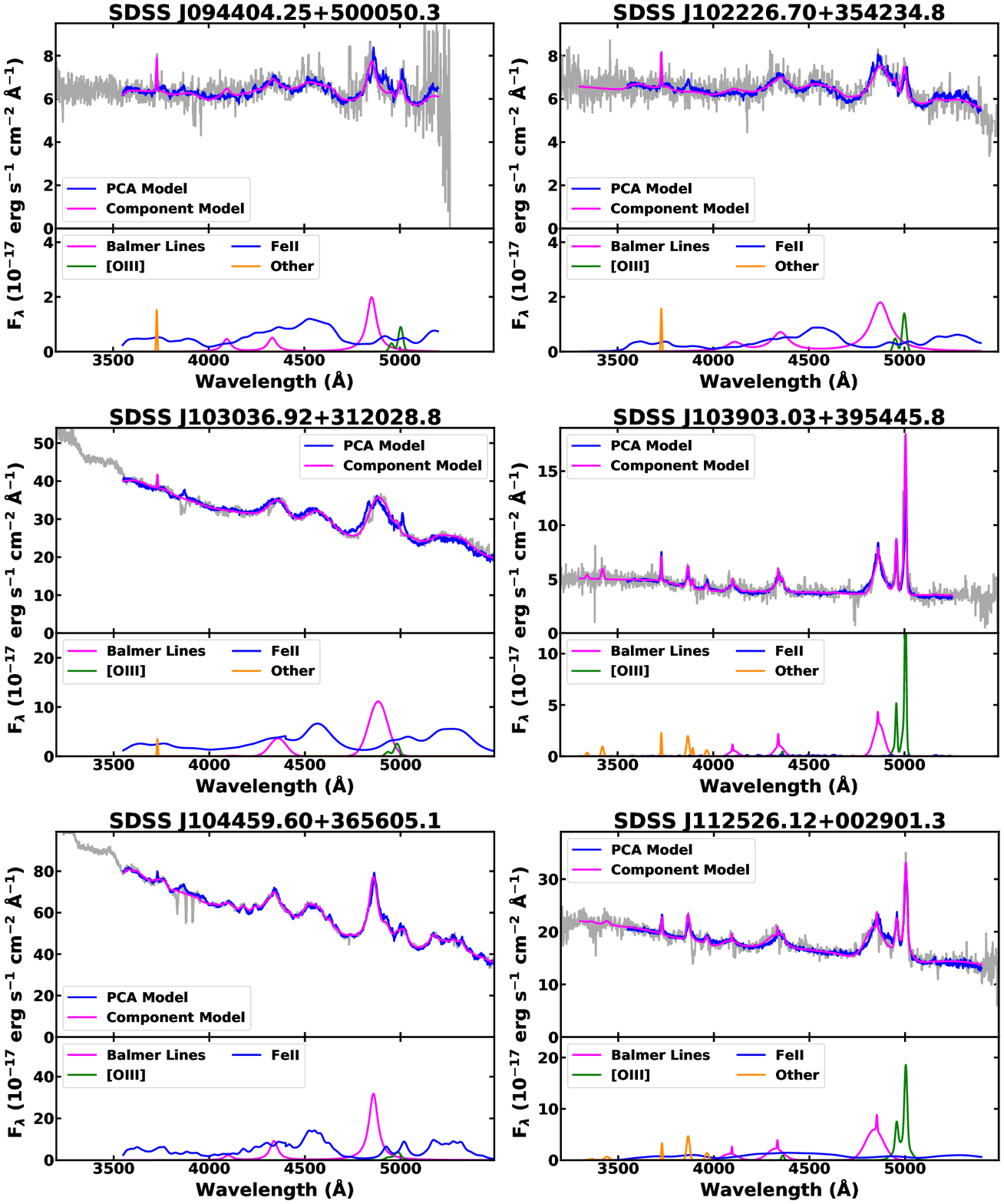}
\caption{Continued.  \label{fig1b}} 
\end{center}
\end{figure*}

\addtocounter{figure}{-1} 
\begin{figure*}[!t]
\epsscale{1.0}
\begin{center}
\includegraphics[width=6.0truein]{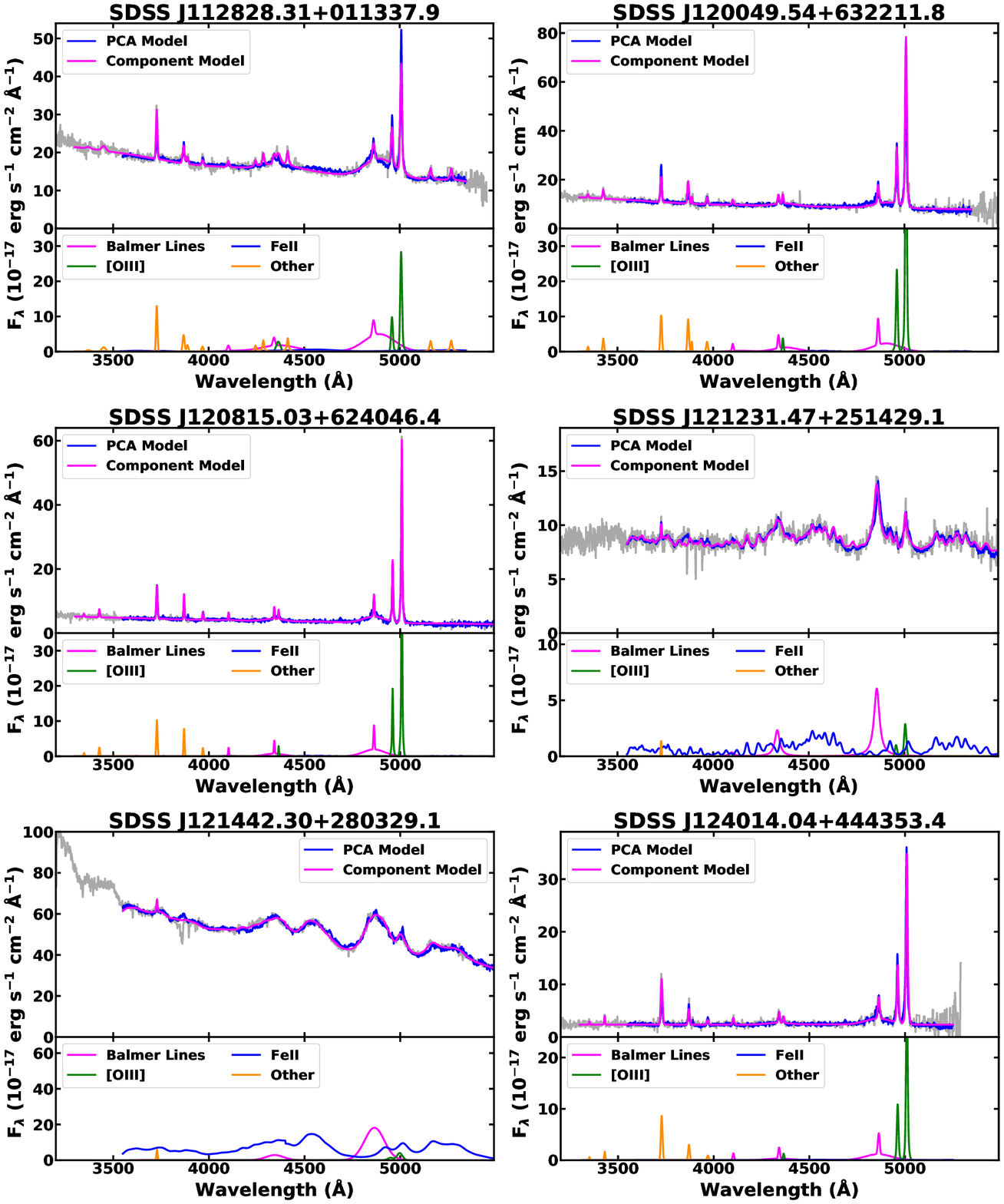}
\caption{Continued.   \label{fig1c}} 
\end{center}
\end{figure*}

\addtocounter{figure}{-1} 
\begin{figure*}[!t]
\epsscale{1.0}
\begin{center}
\includegraphics[width=6.0truein]{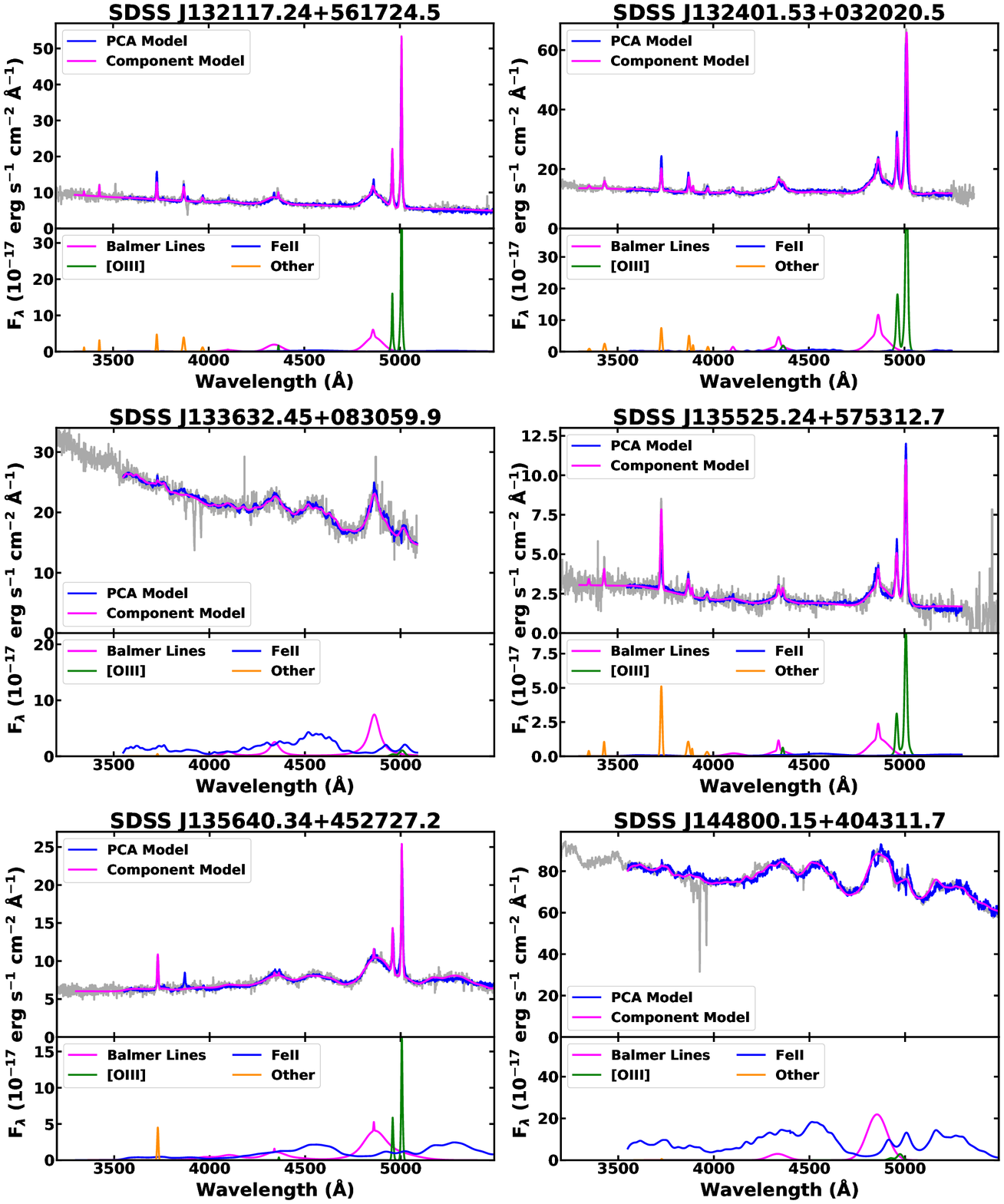}
\caption{Continued.   \label{fig1d}} 
\end{center}
\end{figure*}

\addtocounter{figure}{-1} 
\begin{figure*}[!t]
\epsscale{1.0}
\begin{center}
\includegraphics[width=6.0truein]{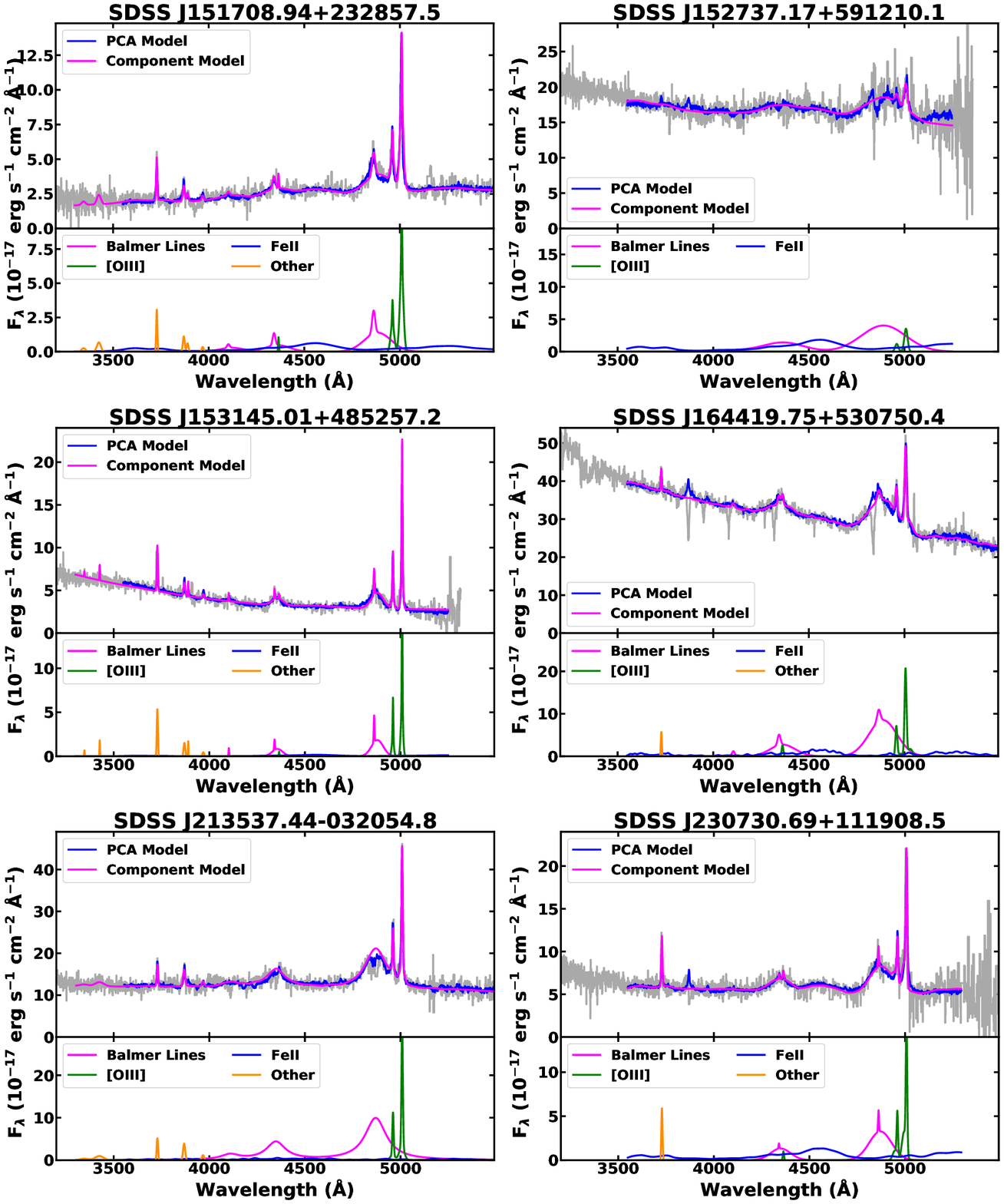}
\caption{Continued.   \label{fig1e}} 
\end{center}
\end{figure*}

\startlongtable
\movetabledown=15mm
\begin{longrotatetable}
\begin{deluxetable*}{LCCCCCCCCCCC}
\tabletypesize{\scriptsize}
\tablecaption{Spectral Fitting Results\label{data}}
\tablehead{
\colhead{SDSS Object Name} & \colhead{H$\beta$ FWHM} &
\colhead{H$\beta$ EW} & \colhead{R$_\mathrm{FeII}$} & \colhead{[\ion{O}{3}]
  EW} & \colhead{[\ion{O}{3}] V$_{50}$} & \colhead{[\ion{O}{3}] W$_{80}$} & 
\colhead{[\ion{O}{3}] Luminosity} & 
\colhead{$\log L_\mathrm{Bol}$} & \colhead{BH Mass} &
\colhead{Edd.\ Ratio} & \colhead{E1 Parameter}\tablenotemark{a}\\
 & \colhead{($\rm km\, s^{-1}$)} & \colhead{(\AA\/)} & 
&  \colhead{(\AA\/)} & \colhead{($\rm km\, s^{-1}$)} & 
\colhead{($\rm km\, s^{-1}$)}  & \colhead{$\rm (erg\, s^{-1}$)} & 
\colhead{$\rm (erg\, s^{-1})$} & \colhead{($\rm M_{\odot}$)}
}
\startdata
015813.56-004635.5 & 6850_{-700}^{+1180} & 93_{-9.8}^{+18.} & 
0.196_{-0.076}^{+0.115} & 106.5_{-3.8}^{+4.2} & 67_{-8}^{+9} & 
530_{-20}^{+20} &  42.87_{-0.02}^{+0.02} & 
45.11\pm 0.11 & 8.57_{-0.11}^{+0.13} & -1.56_{-0.11}^{+0.13} &  -2.45_{-0.42}^{+0.39} \\
025858.17-002827.0 & 9000_{-290}^{+290} & 96_{-3.2}^{+3.1} & 
0.378_{-0.039}^{+0.045} & 4.3_{-0.5}^{+0.5} & -261_{-52}^{+58} & 
1110_{-140}^{+160} & 42.32_{-0.06}^{+0.05} & 
46.36 \pm 0.01 & 8.64_{-0.05}^{+0.04} & -0.38_{-0.05}^{+0.04}  &  1.10_{-0.14}^{+0.16} \\
080248.18+551328.8 & 1560_{-30}^{+30} & 42_{-0.6}^{+0.5} & 
1.67_{-0.045}^{+0.045} & 17.8_{-0.6}^{+0.6} & -111_{-23}^{+21} & 
1670_{-60}^{+60} & 42.82_{-0.01}^{+0.01} & 
46.56\pm 0.009 & 7.57_{-0.04}^{+0.04} & 0.9_{-0.04}^{+0.04} &  1.01_{-0.037}^{+0.039}\\
080957.39+181804.4 & 5400_{-150}^{+140} & 56_{-1.8}^{+1.7} & 
0.743_{-0.062}^{+0.069} & 9_{-1.0}^{+0.9} & -320_{-182}^{+140} & 
2180_{0}^{+0} &  42.99_{-0.05}^{+0.04} & 
46.41\pm 0.02 & 7.97_{-0.04}^{+0.04} & 0.35_{-0.04}^{+0.04} &  0.98_{-0.12}^{+0.13} \\
083522.77+424258.3 & 3250_{-40}^{+40} & 91_{-0.9}^{+0.8} & 
0.696_{-0.018}^{+0.017} & 22.9_{-0.7}^{+0.7} & -274_{-21}^{+21} & 
2100_{-90}^{+80} & 43.18_{-0.01}^{+0.01} & 
46.58\pm 0.01 & 8.33_{-0.04}^{+0.04} & 0.15_{-0.04}^{+0.04} &  0.05\pm
0.04 \\
091658.43+453441.1 & 10390_{-700}^{+1200} & 102_{-7.8}^{+12.} & 
0.121_{-0.107}^{+0.068} & 175.9_{-21.4}^{+20.8} & 36_{-6}^{+4} & 
460_{-40}^{+60} & 43.22_{-0.05}^{+0.05} & 
45.53 \pm 0.03 & 8.9_{-0.07}^{+0.10} & -1.47_{-0.07}^{+0.1} &  -3.33_{-0.67}^{+0.44}\\
094404.25+500050.3 & 3800_{-520}^{+620} & 35_{-4.0}^{+4.7} & 
1.308_{-0.232}^{+0.262} & 4.9_{-1.2}^{+1.3} & -245_{-225}^{+221} & 
1840_{-440}^{+470}  & 42.1_{-0.12}^{+0.1} & 
46.17 \pm 0.02 & 8.46_{-0.15}^{+0.13} & -0.39_{-0.15}^{+0.13} &  2.02_{-0.27}^{+0.32}\\
102226.70+354234.8 & 7360_{-670}^{+740} & 59_{-5.0}^{+5.4} & 
0.538_{-0.10}^{+0.11} & 6.8_{-0.9}^{+0.9} & -537_{-82}^{+75} & 
1630_{-260}^{+290}  & 42.08_{-0.06}^{+0.06} & 
46.05 \pm 0.02 & 8.66_{-0.10}^{+0.09} & -0.7_{-0.10}^{+0.09} &
0.97\pm 0.21\\
103036.92+312028.8 & 7090_{-90}^{+110} & 53_{-0.7}^{+2.0} & 
0.868_{-0.04}^{+0.056} & 4_{-0.4}^{+0.4} & -1581_{-71}^{+76} & 
2130_{-140}^{+140}  & 42.51_{-0.05}^{+0.04} & 
46.64 \pm 0.01 & 8.57\pm -0.04 & -0.02 \pm 0.04 &  1.86_{-0.09}^{+0.12}\\
103903.03+395445.8 & 3930_{-170}^{+140} & 58\pm 2.0 & 
0.075_{-0.078}^{+0.091} & 63_{-5.2}^{+4.9} & -252_{-26}^{+25} & 
1250_{-120}^{+130} & 42.91_{-0.04}^{+0.03} & 
46.18 \pm 0.02 & 8.44_{-0.06}^{+0.05} & -0.35_{-0.06}^{+0.05} &  -2.76_{-0.65}^{+0.72}\\
104459.60+365605.1 & 3460_{-40}^{+30} & 57_{-0.5}^{+0.4} & 
0.832_{-0.017}^{+0.018} & 4.0_{-0.2}^{+0.3} & -1405_{-108}^{+83} & 
2620_{-190}^{+250}  & 42.57\pm 0.03 & 
46.71 \pm 0.009 & 7.82 \pm 0.03 & 0.8 \pm 0.03 &  1.83\pm 0.06 \\
112526.12+002901.3 & 6800_{-130}^{+140} & 46_{-0.8}^{+0.9} & 
0.452_{-0.019}^{+0.013} & 36.8_{-1.0}^{+0.9} & -344_{-22}^{+27} & 
2920_{-170}^{+210}  & 43.27 \pm 0.01 & 
46.28 \pm 0.02 & 8.67 \pm 0.05 & -0.48 \pm 0.05 &  -0.76_{-0.04}^{+0.03}\\
112828.31+011337.9 & 11550_{-360}^{+1160} & 70_{-2.1}^{+8.7} & 
0.125_{-0.056}^{+0.064} & 27.9_{-4.3}^{+4.5} & -42_{-20}^{+38} & 
830_{-90}^{+380} &  43.15\pm 0.07 & 
46.17 \pm 0.02 & 8.29_{-0.05}^{+0.08} & -0.22_{-0.05}^{+0.08}  &  -1.57_{-0.46}^{+0.39} \\
120049.54+632211.8 & 10060_{-680}^{+730} & 47_{-4.2}^{+4.6} & 
0.091_{-0.051}^{+0.062} & 99.3_{-7.6}^{+7.4} & 2_{-10}^{+8} & 
790_{-40}^{+50} & 43.5\pm 0.03 & 
45.89 \pm 0.02 & 8.3 \pm 0.07 & -0.53\pm 0.07 &  -3.03_{-0.56}^{+0.45}\\
120815.03+624046.4 & 8140_{-330}^{+420} & 78_{-2.9}^{+3.4} & 
0.093_{-0.046}^{+0.045} & 144.9_{-10.6}^{+10.7} & -11_{-3}^{+6} & 
590\pm 40 &  43.13\pm 0.03 & 
45.42 \pm 0.04 & 8.31\pm 0.05 & -0.99 \pm 0.05 &  -3.37_{-0.49}^{+0.34}\\
121231.47+251429.1 & 2100\pm 120 & 42.0\pm 1.8 & 
1.102_{-0.103}^{+0.111} & 5.6_{-0.5}^{+0.6} & -240\pm 34 & 
900_{-120}^{+140} & 42.16\pm 0.04 & 
46.34 \pm 0.01 & 7.99_{-0.07}^{+0.06} & 0.25_{-0.07}^{+0.06} &
1.76\pm 0.12 \\
121442.30+280329.1 & 7480\pm 90 & 56\pm 0.7 & 
1.112_{-0.042}^{+0.045} & 3.8\pm 0.3 & -657_{-123}^{+103} & 
2180 &  42.47_{-0.04}^{+0.03} & 
46.47 \pm 0.01 & 8.27\pm 0.03 & 0.1\pm 0.03 &  2.14_{-0.08}^{+0.09} \\
124014.04+444353.4 & 10050_{-1020}^{+1330} & 87_{-9.9}^{+11.0} & 
0.083_{-0.084}^{+0.065} & 162.4_{-23.6}^{+23.0} & 86_{-12}^{+17} & 
790_{-70}^{+100} & 43.25_{-0.07}^{+0.06} & 
45.66 \pm 0.03 & 8.49_{-0.10}^{+0.11} & -0.92_{-0.10}^{+0.11} &  -3.56_{-0.65}^{+0.55} \\
132117.24+561724.5 & 6600_{-150}^{+200} & 74_{-2.1}^{+2.0} & 
0.156_{-0.06}^{+0.015} & 75.1_{-2.1}^{+2.4} & 5_{-5}^{+4} & 
720\pm 20 &  43.09\pm 0.01 & 
45.39 \pm 0.04 & 8.49_{-0.04}^{+0.05} & -1.2_{-0.04}^{+0.05} &  -2.31_{-0.40}^{+0.08}\\
132401.53+032020.5 & 7360_{-310}^{+330} & 60_{-2.1}^{+2.2} & 
0.168_{-0.046}^{+0.055} & 81_{-8.1}^{+8.0} & 165_{-15}^{+12} & 
1200_{-60}^{+80} & 43.61_{-0.05}^{+0.04} & 
46.30 \pm 0.02 & 8.3\pm 0.05 & -0.1\pm 0.05 &  -2.33_{-0.27}^{+0.26}\\
133632.45+083059.9 & 4240_{-330}^{+170} & 48_{-3.2}^{+1.4} &
0.999_{-0.091}^{+0.090} & 2.6\pm 0.5 & 9_{-90}^{+1715} &
2180  & 42.05_{-0.10}^{+0.08} &
46.1\pm 0.01 & 8.21_{-0.08}^{+0.05} & -0.2_{-0.08}^{+0.05} &  2.39_{-0.20}^{+0.22}\\
135525.24+575312.7 & 7560_{-650}^{+680} & 97_{-6.2}^{+7.3} & 
0.24_{-0.239}^{+0.039} & 104.1_{-12.4}^{+8} & -45_{-82}^{+20} & 
1320_{-120}^{+1380} & 42.78_{-0.05}^{+0.03} & 
45.63 \pm 0.03 & 8.47_{-0.09}^{+0.08} & -0.93_{-0.09}^{+0.08}  &  -2.26_{-0.67}^{+0.18}\\
135640.34+452727.2 & 9540_{-900}^{+990} & 140_{-20}^{+24} & 
0.57_{-0.13}^{+0.16} & 27.5\pm 3.4 & -110_{-11}^{+12} & 
640_{-50}^{+56} & 42.67_{-0.06}^{+0.05} & 
46.03 \pm 0.01 & 8.90_{-0.10}^{+0.09} & -0.96_{-0.09}^{+0.10} &  -0.29_{-0.23}^{+0.25}\\
144800.15+404311.7 & 7670\pm 100 & 44_{-0.5}^{+0.6} & 
1.196 \pm 0.033 & 1.6\pm 0.2 & -2084_{-99}^{+120} & 
2180 & 42.49\pm 0.05 & 
46.81 \pm 0.009 & 8.55\pm 0.04 & 0.16\pm 0.04 &  3.00\pm 0.11\\
151708.94+232857.5 & 10080_{-640}^{+1170} & 110_{-6.0}^{+10} & 
0.53_{-0.1}^{+0.05} & 70.5_{-3.0}^{+3.1} & 48_{-12}^{+14} & 
1320_{-90}^{+60} &  42.72\pm 0.02 & 
45.92 \pm 0.02 & 9.02_{-0.07}^{+0.1} & -1.2_{-0.07}^{+0.1} &  -1.24_{-0.18}^{+0.08}\\
152737.17+591210.1 & 16790_{-1080}^{+1170} & 78_{-7.8}^{+8.7} & 
0.32_{-0.062}^{+0.049} & 5.4\pm 1 & -30_{-77}^{+78} & 
1330_{-210}^{+220}  & 42.52\pm 0.08 & 
46.60 \pm 0.01 & 9.28\pm 0.07 & -0.78\pm 0.07 &  0.76_{-0.24}^{+0.21}\\
153145.01+485257.2 & 4460_{-350}^{+340} & 39\pm 3.2 & 
0.239_{-0.201}^{+0.057} & 52.3_{-4.4}^{+3.9} & -10_{-15}^{+11} & 
640_{-60}^{+120} &  42.82_{-0.04}^{+0.03} & 
45.86 \pm 0.02 & 8.67_{-0.09}^{+0.08} & -0.9_{-0.09}^{+0.08} &  -1.62_{-0.65}^{+0.21} \\
164419.75+530750.4 & 10590_{-160}^{+290} & 54_{-0.9}^{+1.3} & 
0.176_{-0.020}^{+0.043} & 12.2_{-0.7}^{+0.5} & -79_{-27}^{+17} & 
1730_{-420}^{+290}  & 42.95_{-0.03}^{+0.02} & 
46.38 \pm 0.009 & 8.97_{-0.04}^{+0.05} & -0.69_{-0.04}^{+0.05} &  -0.51_{-0.11}^{+0.19} \\
213537.44-032054.8 & 8270_{-370}^{+380} & 185_{-5.0}^{+5.2} & 
0.038\pm 0.01 & 33.2_{-1.8}^{+1.7} & 5_{-12}^{+13} & 
870_{-70}^{+110} &  43.06_{-0.03}^{+0.02} & 
46.41 \pm 0.01 & 8.75\pm 0.06 & -0.44\pm 0.06 &  -2.74_{-0.27}^{+0.20}\\
230730.69+111908.5 & 8430_{-280}^{+340} & 102_{-3.2}^{+4.4} & 
0.564_{-0.104}^{+0.120} & 46.7_{-3.7}^{+3.3} & -174_{-31}^{+34} & 
1480_{-170}^{+140}  & 42.92_{-0.04}^{+0.03} & 
46.31 \pm 0.02 & 8.69\pm 0.05 & -0.47\pm 0.05 &  -0.80\pm 0.18 \\
\enddata
\tablenotetext{a}{The $E1$ parameter is defined in
  \S\ref{distributions}.}
\end{deluxetable*}
\end{longrotatetable}

\subsection{Principal Components Analysis}\label{pca}

While spectral decomposition using emission-line profiles and
templates is a widely used and valid approach, model dependence must 
be present.  Decomposition using spectral principal components provides a
complementary and less model dependent analysis.

We created the principal components using the 132 comparison spectra
described in \S\ref{selection}.  We preprocessed the spectra by
dividing by the fitted continuum flux density at 4950\AA\/, and
subtracting the scaled continuum obtained from spectral fitting. This
procedure of subtracting the continuum differs from many applications
of PCA \citep[e.g.,][]{shang03,  paris11} and has the advantage that
the considerable variance present due to variations in continuum slope
does not contaminate the emission-line eigenvectors.  

\begin{figure*}[!t]
\epsscale{1.0}
\begin{center}
\includegraphics[width=4.0truein]{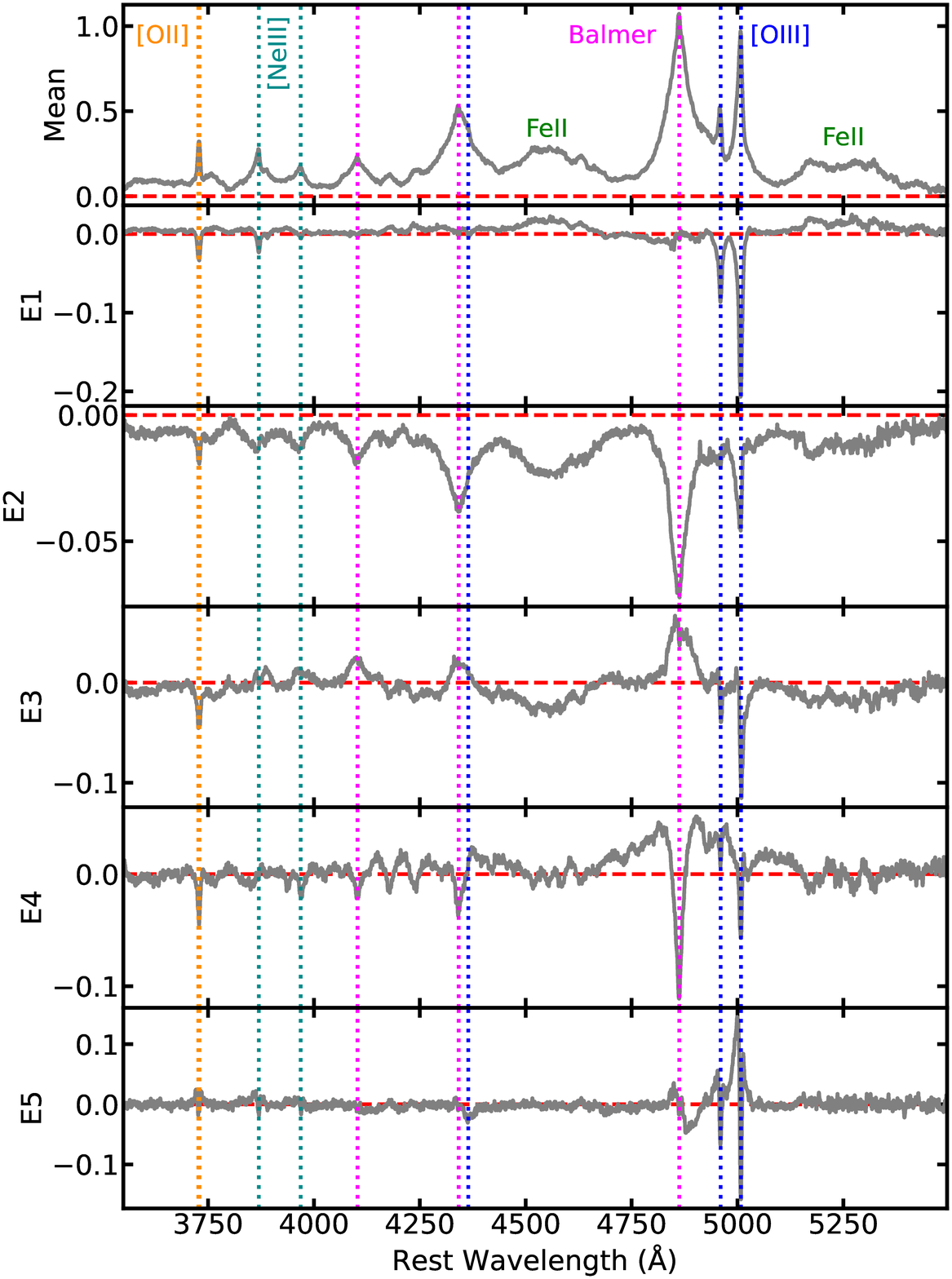}
\caption{The mean spectrum and first five eigenvectors constructed
  from the 132 objects in the comparision sample.  Principal emission
  lines are labeled.    \label{pca_eigen}}
\end{center}
\end{figure*}

We used {\tt EMPCA} \citep{bailey12} to compute the eigenvectors.  
This method has several advantages over the traditional
singular value decomposition method. Principally, the method 
weights the data using the uncertainties, and  missing data are given a
weight of zero.   This is important since PCA methods find the regions of
the data that have the most variance, noise can infiltrate
the eigenvectors that describe the most variance in the data set
\citep[e.g., Fig.\ 5 in][]{bailey12}. The
first and second eigenvectors modeled 29\% and 9.9\% of the variance,
respectively.  We retain the first five eigenvectors to model the variance of
the emission lines; subsequent ones each model less than 1\% of the
variance.  The mean and first five  eigenvectors are shown 
in Fig.~\ref{pca_eigen}. 

The mean spectrum and eigenvectors offer no great surprises.  The
first eigenvector shows the strong anticorrelation between
[\ion{O}{3}] and \ion{Fe}{2}, and is essentially the \citet{bg92} 
first eigenvector.  This construct is historically known as
Eigenvector 1, and is repeatedly found in PCA analysis of optical
rest-frame quasar spectra \citep[e.g.,][]{sulentic00,
  grupe04,ludwig09,wolf20}.    The second 
eigenvector shows strong Balmer lines 
and \ion{Fe}{2} emission.  The fourth eigenvector is interesting
because it includes very narrow \ion{Fe}{2} lines, and therefore
should be strongly negative when modeling the narrow-line Seyfert 1
galaxies in our sample.  It is notable that the [\ion{O}{2}] line and
other very narrow lines are clearly present in the spectra; this is
a consequence of the careful redshift correction.  

We used the mean spectrum and first five eigenvectors to fit the 30
FeLoBAL spectra and the 132 comparison sample spectra. The spectra
were prepared for spectral fitting by being normalized by the same
value used to process the spectra for PCA analysis.  The model is
comprised of a linear combination of the same continuum model used for
each spectrum in \S\ref{optical_modeling}, the mean emission-line
spectrum, and the five eigenvectors, where the normalization of the
mean emission-line spectrum is frozen at a value of 1.  The continuum
normalization was left free to vary, as were the coefficients of the
eigenvectors which serve to modify the shape of the mean emission line
spectrum.  

\section{Distribution Comparisons and Correlations}\label{dist_comp}

In this section, we compare the emission-line properties of the
FeLoBAL quasars  with those of the comparison sample of unabsorbed
quasar using cumulative distribution plots. We then examine
relationships among the optical and global parameters using
Spearman-rank correlations.

\subsection{Distributions}\label{distributions}

 We report the results of
the two-sample Kolmogorov-Smirnov  (KS) test and the two-sample
Anderson-Darling (AD) test in 
Table~\ref{distrib_tab}.  The KS test reliably tests the difference
between two distributions when the difference is large at the median
values, while the AD test is more reliable if the differences lie
toward the maximum or minimum values (i.e., the median can be
the same, and the distributions different at larger and smaller
values)\footnote{E.g.,
  https://asaip.psu.edu/articles/beware-the-kolmogorov-smirnov-test/}. We
also examined the relationships among the optical properties of our 
sample of FeLoBALQs and of the comparison sample.  
The four pairs of columns in the table correspond to several different
cases.  The first pair gives the results for all the data; the
remainder are divided by the $E1$ parameter which is described next.

\movetabledown=0.5in
\begin{longrotatetable}
\begin{deluxetable*}{lCCCCCCCCCCC}
\tabletypesize{\scriptsize}
\tablecaption{Parameter Distributions Comparison\label{distrib_tab}}
\tablehead{
\colhead{Parameter Name} & \multicolumn{6}{c}{FeLoBALQs vs Comparison}
& \multicolumn{2}{c}{FeLoBALQs} \\
\colhead{} & \multicolumn{2}{c}{All (30/132)} &
\multicolumn{2}{c}{E1$<0$ (17/61)} &
 \multicolumn{2}{c}{E1$>0$ (13/71)}  &  \multicolumn{2}{c}{E1$<0$ vs
   E1$>0$\tablenotemark{a}}  \\
\colhead{} & \colhead{KS\tablenotemark{b}} & \colhead{AD\tablenotemark{c}} & \colhead{KS\tablenotemark{b}} & \colhead{AD\tablenotemark{c}} & \colhead{KS\tablenotemark{b}} & \colhead{AD\tablenotemark{c}} 
& \colhead{KS\tablenotemark{b}} & \colhead{AD\tablenotemark{c}} \\
}
\startdata
H$\beta$ FWHM &  {\bf 0.56 / 1.5\times 10^{-7}} & {\bf 18.0 / <0.001} &
{\bf 0.65 / 6.3\times 10^{-6}} &  {\bf 13.3 / < 0.001} & 
{\bf 0.42 /  0.029} & {\bf 4.25 / 6.5\times 10{-3}} &  
0.43 / 0.09 & {\bf 3.3 / 0.014} \\
H$\beta$ EW & 0.25 / 0.088 & 0.99 / 0.13 &
0.24 / 0.37 &  0.19 / >0.25 &
0.37 / 0.068 & 1.1 / 0.12 &
{\bf 0.48 / 0.050} & {\bf 2.65 / 0.027} \\
R$_\mathrm{FeII}$ & {\bf 0.35 / 3.6\times 10^{-3}} & {\bf 4.5 /  5.4\times 10^{-3}} 
& {\bf 0.60 / 4.9\times 10^{-5}} & {\bf 7.3 / <0.001} & 
0.14 / 0.97 & -0.83 / >0.25 & 
{\bf 0.77 / 6.8 \times 10^{-5}} & {\bf 11.8 / <0.001} \\
$[$\ion{O}{3}$]$ EW & {\bf 0.28 / 0.034} & {\bf 6.5 / 1.0\times 10^{-3}} & 
{\bf 0.47 / 3.0\times 10^{-3}} & {\bf 8.0 / <0.001} & 
{\bf 0.61 / 2.1\times 10^{-4}} & {\bf 8.3 / <0.001} & 
{\bf 0.94 / 2.3\times 10^{-7}} & {\bf 14.1 / <0.001} \\
$[$\ion{O}{3}$]$ V$_{50}$ (km s$^{-1}$) & 0.25 / 0.084 & 1.8 / 0.056 & 
{\bf 0.44 / 6.8\times 10^{-3}} & {\bf 5.3 / 2.7\times 10^{-3}} & 
0.29 / 0.26 & 1.25 / 0.10 & 
{\bf 0.67 / 1.2\times 10^{-3}} & {\bf 7.3 / <0.001} \\
$[$\ion{O}{3}$]$ W$_{80}$ (km s$^{-2}$) & 0.15 / 0.61 & -0.50 / >0.25 & 
{\bf 0.40 / 0.021} & {\bf 2.4 / 0.032} & 
{\bf 0.40 / 0.040} & {\bf 2.0 / 0.050} & 
{\bf 0.67 / 1.2\times 10^{-3}} & {\bf 7.5 / <0.001} \\
SPCA E1 & 
{\bf 0.33 / 7.2\times 10^{-3}} & {\bf 5.0 / 3.5\times 10^{-3}} & 
{\bf 0.53 / 6.0\times 10^{-4}} & {\bf 9.8 / <0.001} &
0.38 / 0.062 & {\bf 2.0 / 0.048} & 
{\bf 1.0 / 1.7 \times 10^{-8}} & {\bf 14.8 / <0.001} \\
SPCA E2 & 
{\bf 0.40 / 4.7\times 10^{-4}} & {\bf 5.1 / 3.1 \times 10^{-3}} & 
0.21 / 0.53 & -0.64 / > 0.25 & 
{\bf 0.66 / 4.8\times 10^{-5}} & {\bf 8.4 / <0.001} &
{\bf 0.48 / 0.050} & {\bf 2.7 /0.026} \\
SPCA E3 & 
{\bf 0.37 / 1.5\times 10^{-3}} & {\bf 10.3 / <0.001} &
{\bf 0.63 / 1.3\times 10^{-5}} & {\bf 10.9 / <0.001} & 
{\bf 0.39 / 0.048} & 1.3 / 0.09 & 
{\bf 0.69 / 8.0\times 10^{-4}} & {\bf 5.5 / 2.3\times 10^{-3}} \\
SPCA E4 & 
0.14 / 0.70 & -0.14 / >0.25 & 
{\bf 0.38 /0.032} & {\bf 3.0 / 0.020} & 
0.35 / 0.10 & 1.2 / 0.10 & 
{\bf 0.55 / 0.014} & {\bf 3.6 / 0.011} \\
\hline
E1 Parameter & 
{\bf 0.34 / 4.7\times 10^{-3}} & {\bf 5.7 / 1.9 \times 10^{-3}} & 
{\bf 0.57 / 1.6\times 10^{-4}} & {\bf 9.0 / < 0.001} & 
{\bf 0.49 / 6.9\times 10^{-3}} & {\bf 5.0 / 3.4\times 10^{-3}} & 
{\bf 1.0 / 1.7\times 10^{-8}} & {\bf 14.8 / < 0.001} \\
H$\beta$ FWHM Deviation & 
{\bf 0.33 / 7.9\times 10^{-3}} & {\bf 6.8 / <0.001} & 
{\bf 0.36 / 0.050} & {\bf 3.0 / 0.019} & 
{\bf 0.42 / 0.026} & {\bf 3.5 / 0.012} & 
0.34 / 0.29 & 0.26 / >0.25 \\
\hline
[\ion{O}{3}] Luminosity & 
0.17 / 0.41 & -0.07 / >0.25 &
0.25 / 0.33 & 0.48 / 0.21 &
0.25 /0.43 & -0.58 / > 0.25 &
{\bf 0.77 / 6.8\times 10^{-5}} & {\bf 9.0 / <0.001} \\
M$_\mathrm{BH}$ & 
0.12 / 0.79 & -0.67 / >0.25 &
0.31 / 0.12 & 0.91 / 0.14 &
0.27 / 0.32 & 0.88 / 0.14 &
{\bf 0.59 / 6.0\times 10^{-3}} & {\bf 5.5 / \bf 2.2\times 10^{-3}} \\
L$_\mathrm{Bol}$ & 
0.11 / 0.91 & -0.72 / >0.25 &
0.31 / 0.12 & 1.65 / 0.068 & 
{\bf 0.47 / 9.0\times 10^{-3}} & {\bf 3.6 / 0.011} &
{\bf 0.65 / 1.7\times 10^{-3}} & {\bf 8.5 / <0.001} \\
L$_\mathrm{Bol}$/L$_\mathrm{Edd}$ & 
0.20 / 0.23 & {\bf 3.1 / 0.018} & 
0.35 / 0.06 & {\bf 3.5 / 0.013} & 
{\bf 0.50 / 4.2\times 10^{-3}} & {\bf 6.9 / <0.001} &
{\bf 0.72 / 2.9\times 10^{-4}} & {\bf 9.7 / <0.001} \\
R$_\mathrm{2800}$ [pc] & 0.16 / 0.48 & 0.16 / >0.25 & 
0.30 / 0.16 & 0.43 / 0.22 &
0.17 / 0.86 & -0.76 / >0.25 & 
0.18 / 0.93 & -0.77 / >0.25   \\
\enddata
\tablenotetext{a}{The optical data from the $E<0$ ($E>0$) FeLoBALQ
  subsamples  include 17 (13) objects.  The {\it SimBAL} data from the $E1<0$
  ($E1>0$) subsamples include 19 (17) measurements because several
  objects have more than one outflow component.}
\tablenotetext{b}{The Kolmogorov-Smirnov Two-sample test.  Each entry
  has two numbers:  the first is the value of the statistic, and the
  second is the probability that the two samples arise from the same
  parent sample.  Bold type indicates entries that yield $p<0.05$.}
\tablenotetext{c}{The Anderson-Darling Two-sample test.  Each entry
  has two numbers:  the first is the value of the statistic, and the
  second is the probability that the two samples arise from the same
  parent sample.  Note that the implementation
  used does not compute a probably   larger than 0.25 or smaller than
  0.001. Bold type indicates entries that yield $p<0.05$.}
\end{deluxetable*}
\end{longrotatetable}

\begin{figure*}[!t]
\epsscale{1.0}
\begin{center}
\includegraphics[width=6.5truein]{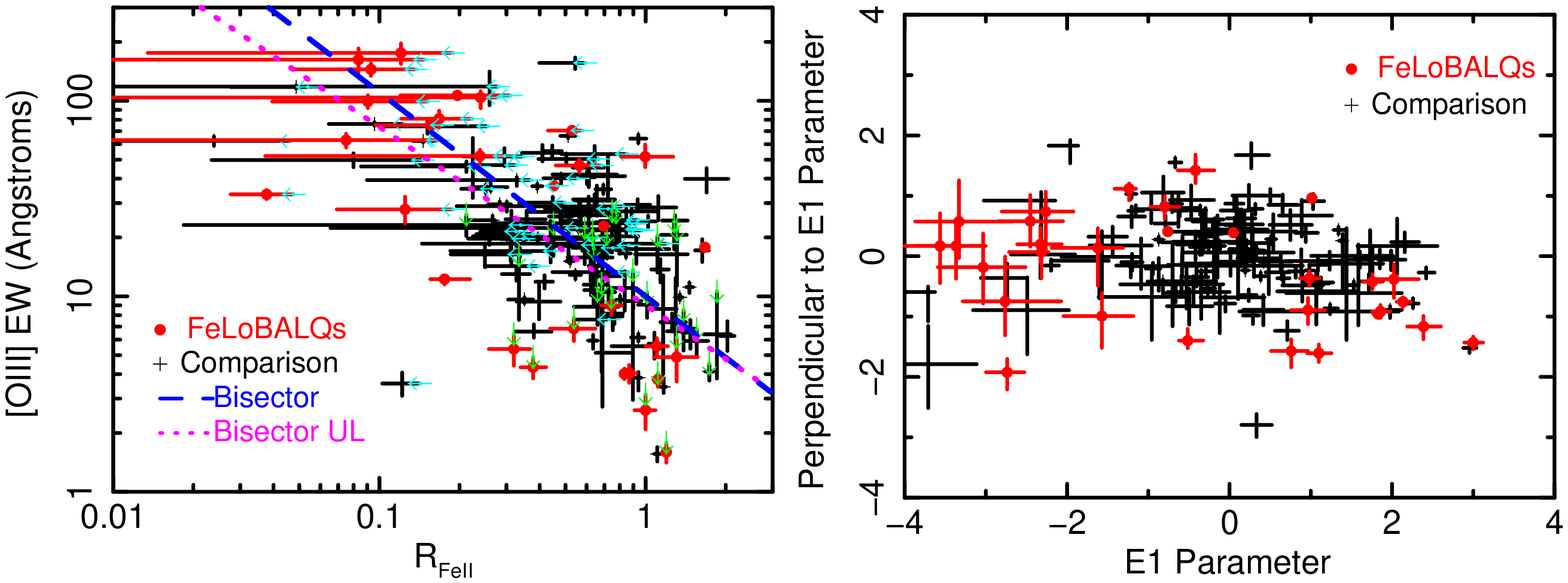}
\caption{The construction of the E1 parameter.  {\it Left:} The
  anticorrelation of the ratio of the \ion{Fe}{2} and H$\beta$ fluxes
  with the [\ion{O}{3}] equivalent width.  Arrows mark the
  measurements deemed statistically insignificant using the F test
  (\S\ref{optical_modeling}).  The new coordinate system was defined
  by the bisector 
  slope computed from the normalized and scaled log values of these
  parameters.  The relationship is given by
  $E1=0.679(\log_{10}(R_\mathrm{FeII})+0.270)-0.735(\log_{10}([OIII]EW)-1.288)$.
  {\it Right:} The data after   rotation to   the new coordinate system.
  The bisector line now lies along the $x$ axis and thereby defines
  the $E1$ parameter.   \label{e1_param}}   
\end{center}
\end{figure*}

The principal components analysis revealed that the dominant variance
in the emission line properties arises from the anticorrelation
between the [\ion{O}{3}] and the \ion{Fe}{2} (\S\ref{pca}).  As noted above, this
anticorrelation is common among samples of optical spectra of quasars,
and also appears among the UV emission lines and in other properties
of quasars
\citep[e.g.,][]{francis92,bg92,brotherton94,corbin96,wills99,shang03,grupe04,yip04,wang06,ludwig09,shen16}.
In this sample, we parameterize these 
properties using 
the [\ion{O}{3}] equivalent width and the ratio of \ion{Fe}{2} to
H$\beta$ \citep[hereafter referred to as $R_\mathrm{FeII}$;
  e.g.,][]{shen_ho_14}.   The anticorrelation for the FeLoBALQs and 
comparison sample is seen in the left panel of Fig.~\ref{e1_param}.  

We created a summary statistic parameter that describes the
anticorrelation.     We computed the base-10 logarithm of
$R_{FeII}$ and [\ion{O}{3}].   {For convenience,} we computed the
mean and standard deviation of both quantities from the comparison
sample, and used those values to scale and normalized both samples.
We then  used the 
bivariate correlated errors and intrinsic scatter (BCES)
method\footnote{https://github.com/rsnemmen/BCES} \citep{ab96} to
obtain the bisector of the two parameters.  This 
method takes into account the uncertainty on both parameters.  The
bisector is more appropriate than a regression, since neither
parameter can be identified as the dependent or independent variable.
After scaling and normalization, the slope of the bisector is close to
$-1$, as expected.   

{ An alternative method for obtaining our summary statistic parameter
takes into account the fact that [\ion{O}{3}] and \ion{Fe}{2} emission
was not found to be statistically necessary using the F test
(\S\ref{optical_modeling}).  We evaluated the bisector using the
methodology of \citet{isobe90} taking into account the statistically
unnecessary values as upper limits using the likelihood given by
\citet{sawicki12}.  The relationship is shown in Fig.~\ref{e1_param}.
As expected, the bisector line is shifted slightly toward lower values
of $R_\mathrm{FeII}$ and [\ion{O}{3}] equivalent width.  Noting that
the summary statistic is defined for convenience and does 
not represent a physical relationship, we proceeded with the original
bisector because it appears more representative of the data.  }

We then rotated the coordinate system so that the long axis of the
bisector is horizontal.  The result is shown in the right panel of
Fig.~\ref{e1_param}.  We take the rotated X-axis value as our desired
summary statistic, hereafter referred to as the ``E1
parameter''. Note that this parameter is closely related to the SPCA1
fit coefficients. As we will discuss in \S\ref{twopop}, the FeLoBALQs
consist of two distinct  populations simply divided at $E1=0$.  In
Table~\ref{distrib_tab} we compare the FeLoBALQs with $E1<0$ and
$E1>0$ with their unabsorbed sample counterparts, and with one another also.

\subsection{FeLoBALQs versus Unabsorbed Objects}  \label{targvscomp}

\subsubsection{Optical Properties} \label{optical}

\begin{figure*}[!t]
\epsscale{1.0}
\begin{center}
\includegraphics[width=6.0truein]{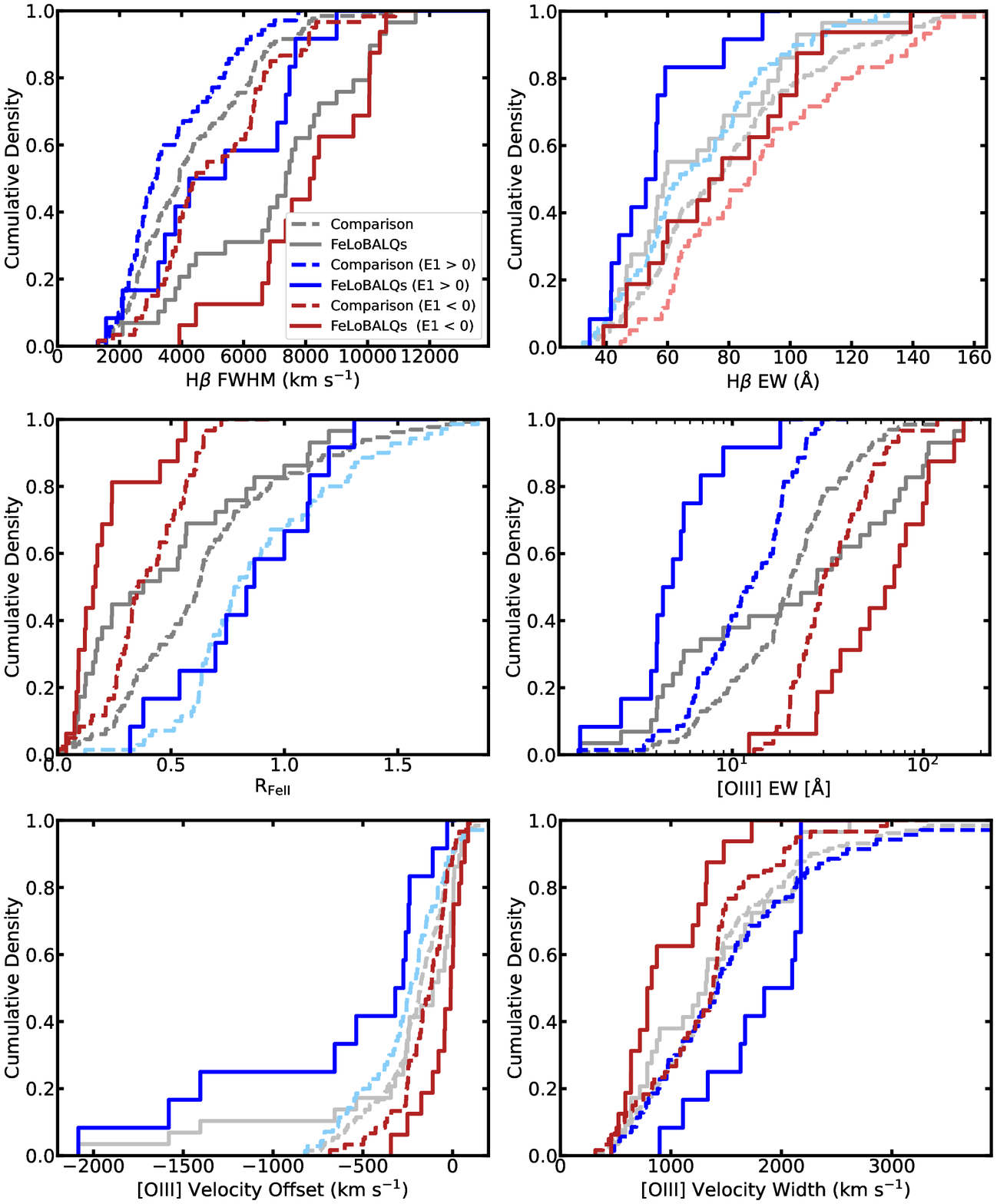}
\caption{Cumulative distribution plots of the optical emission-line
  parameters.  The data are sampled in three ways:  the grey lines
  show the full FeLoBALQ and comparison samples, and the blue (red)
  lines sho  the FeLoBALQ and comparison samples for E1 parameter $E1
  > 0$ ($E1<0$); the E1 parameter is introduced in \S\ref{dist_comp}
  and discussed in \S\ref{twopop}.  
  The Kolmogorov-Smirnov and Anderson-Darling statistics 
  for four different comparisons are given in
  Table~\ref{distrib_tab}.   Distributions that are significantly
  different ($p<0.05$) are shown in dark red, dark blue, or dark grey,
  while distributions that are not significantly different are shown
  in pale colors.  The full  sample distributions are
  statistically   significantly different for  H$\beta$ FWHM,
  R$_\mathrm{FeII}$, and   [\ion{O}{3}] EW.  The FeLoBALQs  partitioned
  by the E1 parameter are statically significantly different for all
  properties.  
   \label{plot_comp_1}} 
\end{center}
\end{figure*}

The H$\beta$ emission line was parameterized using the FWHM and the
equivalent width. The cumulative distributions of the H$\beta$ FWHM
are plotted in Fig.~\ref{plot_comp_1}.  The cumulative distributions
for the FeLoBALQs and comparison sample are profoundly different (see
Table~\ref{distrib_tab}), with the FeLoBALQs showing consistently
larger FWHM.     There is a tendency for the FeLoBALQs to
have smaller H$\beta$ equivalent widths, but it is not statistically
significant.

There is a statistically significant difference in the
R$_\mathrm{FeII}$  between the comparison sample and the FeLoBALQs for
the $E1<0$ subsample.  There are excess low R$_\mathrm{FeII}$ objects 
among the FeLoBALQs, and a hint of excess high R$_\mathrm{FeII}$ objects
as well.  We will return to this point in \S\ref{twopop}.

We used the [\ion{O}{3}] equivalent width to quantify the strength of the
[\ion{O}{3}] emission line, and  $v_{50}$  and $w_{80}$ (defined in
\S\ref{optical_modeling}) to parameterize the profile shape.   The
distributions of the [\ion{O}{3}] equivalent width are 
statistically significantly different; note that this is an excellent
example of a pair of distributions that are found to be less
significantly different using the KS test ($p=0.034$)
compared with the AD test ($p<0.001$; the software that we used does not
compute probabilities less than this value) due to the fact that the
median of the distributions is not very different, but the ends are.
Specifically, there are many more FeLoBALQs with much larger and much
smaller [\ion{O}{3}] equivalent widths than the comparison
sample. There are no statistical differences between the FeLoBALQs and
the comparison sample for  $v_{50}$ and $w_{80}$. 

\subsubsection{The SPCA Eigenvector Coefficients}\label{spca_coeff}

\begin{figure*}[!t]
\epsscale{1.0}
\begin{center}
\includegraphics[width=6.5truein]{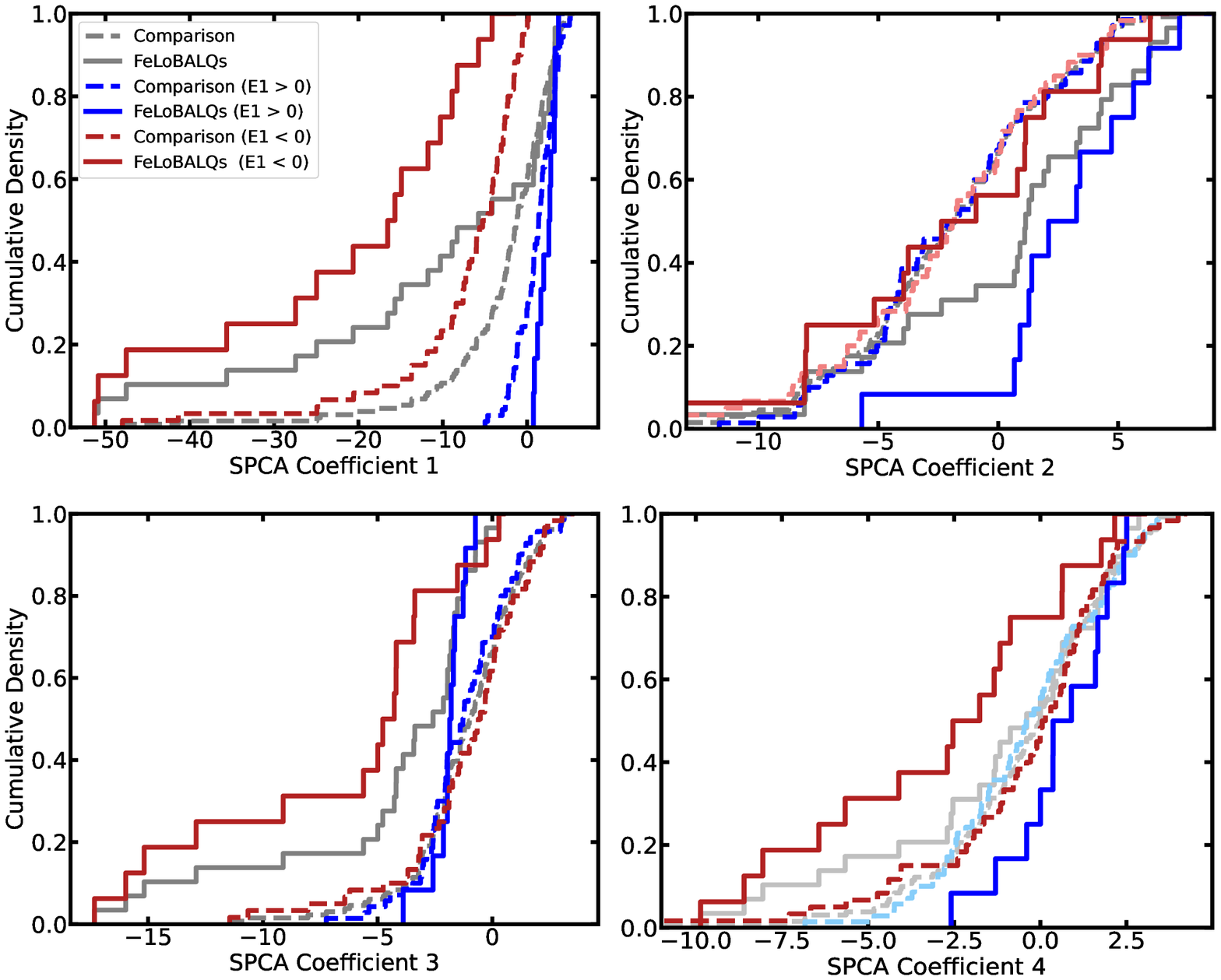}
\caption{Cumulative distribution plots of the SPCA fit coefficients.
  The colors and line styles are the same as in
  Fig.~\ref{plot_comp_1}, and the Kolmogorov-Smirnov and
  Anderson-Darling statistics   for four different comparisons are
  given in Table~\ref{distrib_tab}.   The full  sample distributions
  are statistically   significantly different for  the coefficients
  for the first three eigenvectors and are different for all four
  eigenvectors for the two populations of FeLoBALQs. 
   \label{plot_comp_2}} 
\end{center}
\end{figure*}

We compared the fit coefficients for the model using first five  SPCA
eigenvectors (\S\ref{pca}) in Fig.~\ref{plot_comp_2}, and the KS 
and AD statistics are given in Table~\ref{distrib_tab}.  We found
that the distributions for the FeLoBALQs differs from those of the
comparison sample for the first three coefficients.

\subsubsection{Global Properties}\label{global_sec}

\begin{figure*}[!t]
\epsscale{1.0}
\begin{center}
\includegraphics[width=5.5truein]{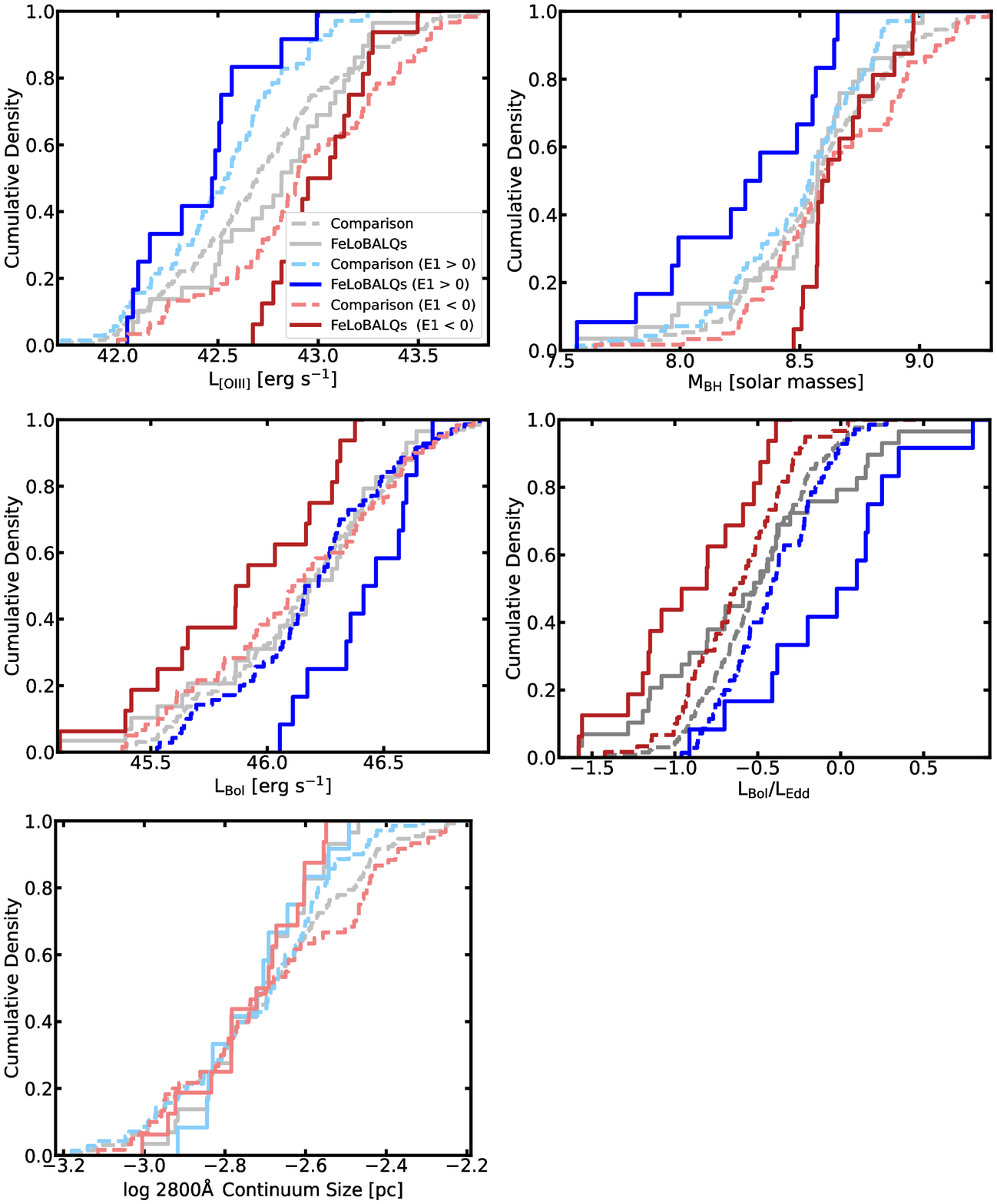}
\caption{Cumulative distribution plots of the global properties: the
  integrated [\ion{O}{3}] luminosity, the bolometric luminosity, the
black hole mass (computed as described in \S\ref{optical_modeling}), the
Eddington ratio L$_\mathrm{Bol}$/L$_\mathrm{Edd}$, and the size of the
2800\AA\/ continuum emission region.  The distributions
are consistent between the FeLoBALQs and comparison sample for most of
these properties; the exception is the Anderson-Darling detection of a
difference in the Eddington ratio.  However, the two E1-parameter
classes of FeLoBALQs are significantly different for the first four
parameters.
   \label{cumulative_global}} 
\end{center}
\end{figure*}

We investigated the distributions of several global properties of the
quasars including the integrated [\ion{O}{3}] luminosity, the
bolometric luminosity, the black hole mass (computed as described in
\S\ref{optical_modeling}), the Eddington ratio
L$_\mathrm{Bol}$/L$_\mathrm{Edd}$, and the size of the 2800\AA\/
continuum emission region.   The distributions are shown in
Fig.~\ref{cumulative_global} and the KS and AD statistics are given in
Table~\ref{distrib_tab}.  Most of  these properties are statistically
consistent between the FeLoBALQs and the comparison sample objects. The
exception is the Anderson-Darling test for the Eddington ratio; the
FeLoBALQs have an excess of objects at high and low
L$_\mathrm{Bol}$/L$_\mathrm{Edd}$ values.

\subsection{Evidence for Two Populations of FeLoBAL
  Quasars}\label{twopop}

\begin{figure*}[!t]
\epsscale{1.0}
\begin{center}
\includegraphics[width=6.5truein]{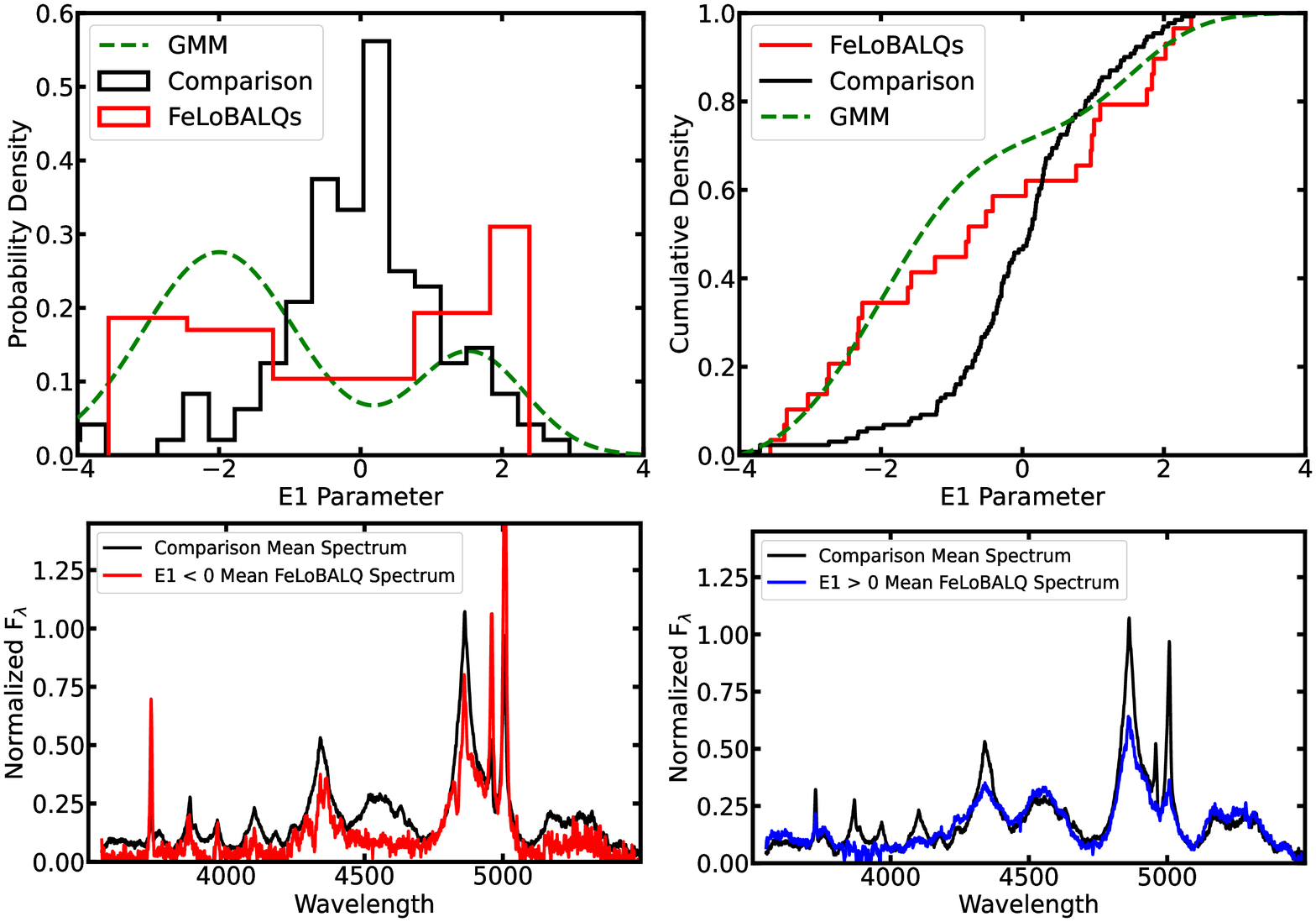}
\caption{The distribution of the E1 parameter for the FeLoBALQs is
  different than for the comparison sample.  {\it Top:}  The histogram
  of the E1 parameter (left) and the cumulative distribution (right).
  Both plots show that there is an excess of FeLoBALQs at both low and
  high E1 parameter. The dashed green line shows the two-component
  Gaussian Mixture Model for the FeLoBAL quasars.  {\it Bottom:} The
  weighted mean spectra of the 
  FeLoBALQs with $E1$ parameter less than zero (left) and greater than
  zero (right), both compared with the weighted mean spectrum from
  the comparison sample.  The objects with $E1 <0$ are characterized
  by stronger [\ion{O}{3}] and weaker \ion{Fe}{2} than the mean
  comparison-sample spectrum, while objects with $E1>0$ are
  characterized  by weaker [\ion{O}{3}] and weaker Balmer lines than
  the mean   comparison-sample   spectrum.   
 \label{hist_ave_spec}}  
\end{center}
\end{figure*}

Examination of the right panel of Fig.~\ref{e1_param} reveals that the
FeLoBALQs are not distributed evenly in the E1 parameter.  While
the comparison sample objects distribution peaks at $E1\sim 0$,
declining toward lower and higher values, the FeLoBALQ $E1$ values
straddle both sides of the center.  The Kolmorogov-Smirnov 
(Anderson-Darling) tests yield a probability of 0.46\% (0.19\%) that
the two samples were drawn from the same parent distribution.  The
differences in the distribution are visualized in the histogram and
cumulative distributions shown in the top panel of
Fig.~\ref{hist_ave_spec}.   

{ The FeLoBAL $E1$ parameter distribution appears to be bimodal,
  but statistical tests do not establish bimodality.  We first used
  the \citet{hartigan85} dip test. 
This test could not reject a unimodal distribution ($p=0.22$).}  We
also used the
\citet{mg10} {\tt GMM}
method\footnote{http://www-personal.umich.edu/$\sim$ognedin/gmm/} \citep{mg10}.
The method fits a two-component gaussian mixture model (GMM) to the
data and compares the result to a one-component GMM model using a
likelihood ratio.  A bootstrap method was used to gauge the robustness
of the result.  The $E1$ parameter from the FeLoBAL quasars yielded a
probability of $p=0.045$ that the distribution is unimodal rather than
bimodal.  The best-fitting two-component
GMM\footnote{https://scikit-learn.org/stable/modules/mixture.html} is
seen as the dashed green line in the top left panel of
Fig.~\ref{hist_ave_spec}.  The Gaussian parameters were $\mu_1=1.54$,
$\mu_2=-2.0$, $\sigma_1=0.78$,   $\sigma_2=1.09$, $N_1=0.42$,
$N_2=0.58$.  Note the significant limitation of this method: it
assumes that the two distributions are Gaussian, and there is no
reason to believe that they should be.   Indeed, the cumulative
distribution functions shown in the top right panel of
Fig.~\ref{hist_ave_spec} indicate that the two-Gaussian approximation is
only an approximate representation of the $E1$ parameter from the
FeLoBAL quasars.  

We divided the samples in two, depending on the sign of the $E1$
parameter.  Note that $E1=0$ is a representative dividing point
because the $\log$ [\ion{O}{3}] equivalent width and $R_\mathrm{FeII}$
were normalized and scaled before $E1$ was computed.  There are 17
(61) and 13 (71) FeLoBALQs (comparison sample 
objects) with E1 parameter less than and greater than zero,
respectively.  The weighted mean spectra for  these two groups
compared with the mean comparison-sample spectrum are shown in 
the lower panel of Fig.~\ref{hist_ave_spec}.  The differences are
notable.  The  FeLoBALQ objects 
 with E1 parameter  less than zero are characterized by stronger
 [\ion{O}{3}] and weaker  \ion{Fe}{2} than the comparison sample
 average spectrum, while the  objects with E1 parameter greater than
 zero have similar \ion{Fe}{2}  emission, weaker Balmer lines, and
 weaker [\ion{O}{3}].  It is also  notable that the $E1 < 0$ mean
 spectrum shows Balmer absorption, seen in the notches in the mean
 spectrum near 4830 and 4330\AA\/,  while the $E1 > 0$ mean spectrum
 shows \ion{Ca}{2} and \ion{He}{1}*  absorption, observed in the scrum
 of absorption lines near 3900 \AA\/.  These different lines point to
 differences in the physical conditions of the absorbing outflow
 \citep{choi22}.  We note that because  these are weighted
average spectra,  they reflect the properties  of the objects
with the best  signal-to-noise spectra.

In \S\ref{dist_comp} we compared the distributions of parameters of
the FeLoBALQs with those of the comparison sample.  Here we perform
the same comparison, but divide the two samples by the $E1$ parameter.
We do this because if the FeLoBALQs are
comprised of two classes, then they may differ in different ways from
the similarly selected comparison sample; we do see this behavior in
the parameter comparison below.  We also compare the two groups of
FeLoBALQs because significant differences indicate a defining or
correlated characteristic property of the two groups.  

\begin{figure*}[!t]
\epsscale{1.0}
\begin{center}
\includegraphics[width=4.5truein]{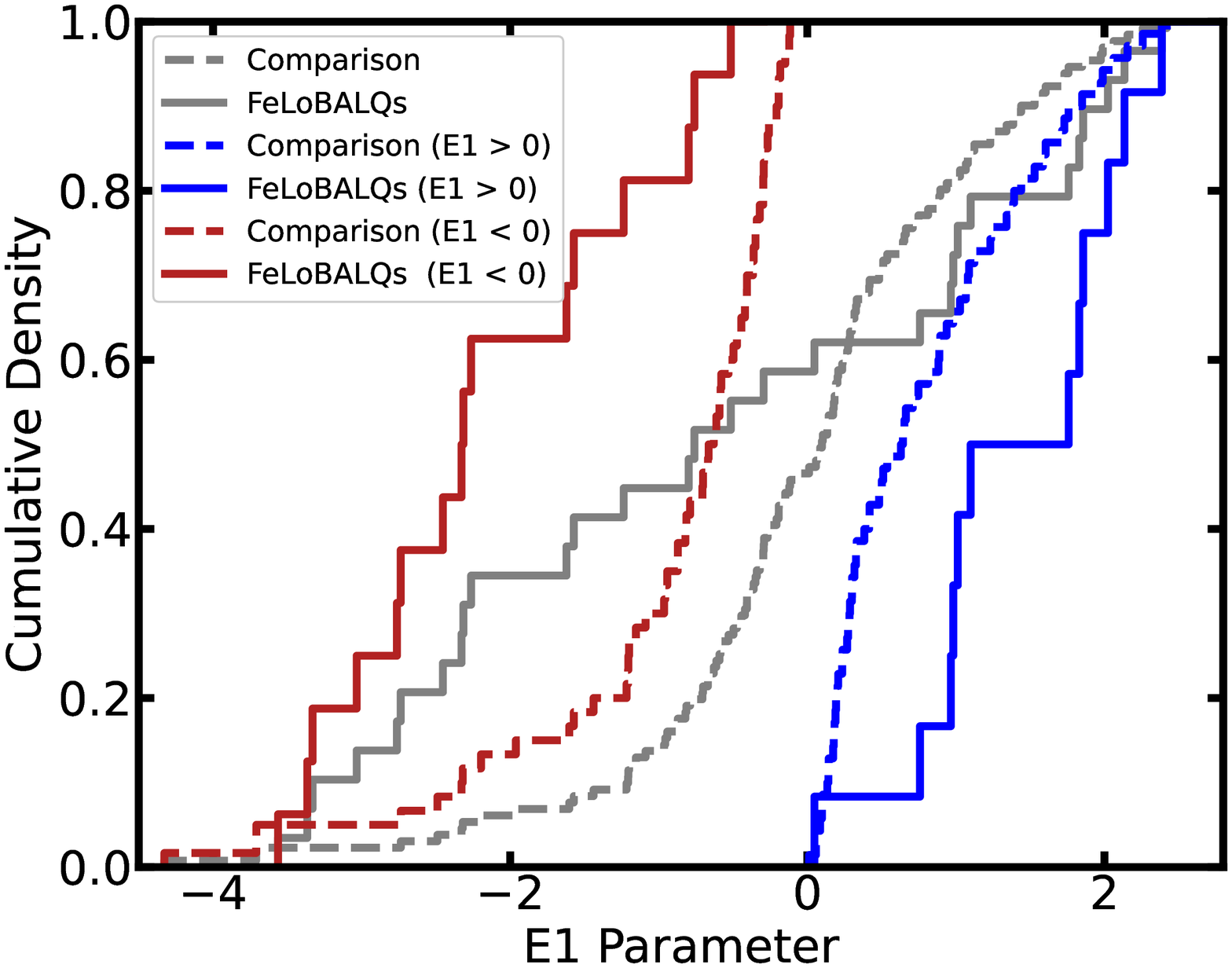}
\caption{The cumulative distributions of the E1 parameter.  The FeLoBAL
  quasars and comparison sample show dramatically different
  distributions, both for the whole sample and for the samples divided
  by positive and negative $E1$.    \label{plot_comp_3}}  
\end{center}
\end{figure*}

The cumulative distribution of the $E1$ parameter is shown in
Fig.~\ref{plot_comp_3}.   Blue (red) lines identify the $E1
> 0$ ($E1 < 0$) objects, with solid (dashed) lines
denoting the FeLoBALQs (comparison) samples.
As expected, dividing the data
by $E1$ results in an enormous difference in the $E1$ parameter
distribution.  

There are significant difference between the FeLoBAL classes among all
of the emission-line properties (Fig.~\ref{plot_comp_1}).  Some of
these are expected; for example,  R$_\mathrm{FeII}$ and [\ion{O}{3}]
EW were used to define the $E1$ parameter.  While as a group the
[\ion{O}{3}] kinematic properties (velocity offset $v_\mathrm{50}$ and
$w_\mathrm{80}$) are consistent between the FeLoBALQs and comparison
objects, they are significantly different between the $E1<0$ and
$E1>0$ FeLoBALQs.  The $E1>0$ ($E1<0$) FeLoBALQs are characterized by
larger (smaller) velocity offsets and broader (narrower) lines. 

The differences between the SPCA coefficients among the different
classes is profound.  Some dependences are expected because E1 and
SPCA coefficient 1 both measure the inverse relationship between the
\ion{Fe}{2} and [\ion{O}{3}] emission.  Other
dependencies are more subtle.  For example, for $E1>0$ FeLoBALQs, SPCA
Coefficient 2 is almost never negative.  This result means that this
eigenvector, characterized by a negative-flux narrow H$\beta$ line,
serves to broaden and reduce the flux of the H$\beta$ line in this
subclass.     It is also
interesting that the coefficients of the first eigenvector are
sufficient to distinguish between the $E1$ subgroups of the comparison
sample; the remaining three are identical for the $E1$ subgroups for
the comparison sample, but not for the FeLoBALQs.  This result
demonstrates the tight relationship between the $E1$ parameter and
SPCA1.  

\begin{figure*}[!t]
\epsscale{1.0}
\begin{center}
\includegraphics[width=4.5truein]{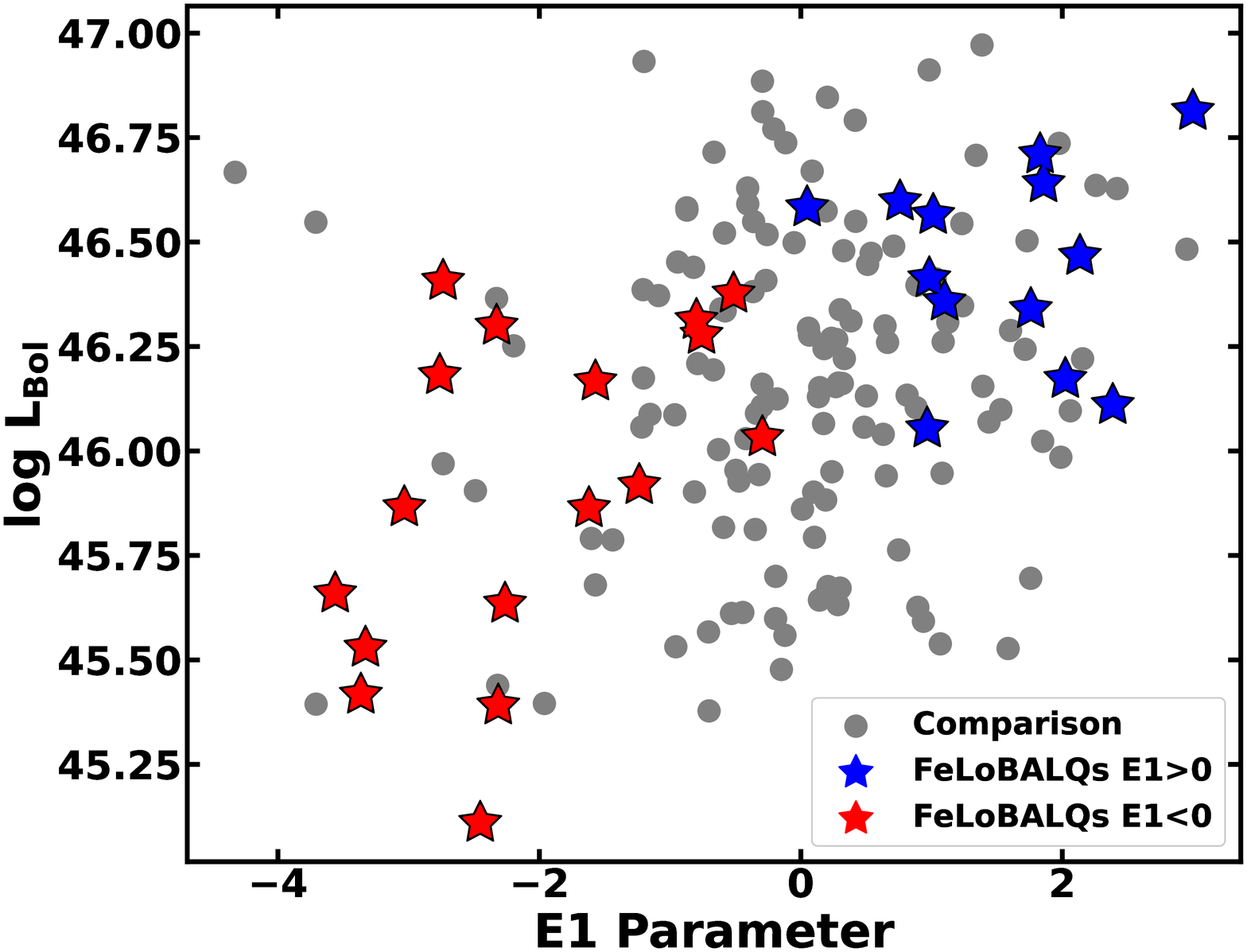}
\caption{The bolometric luminosity as a function of the $E1$
  parameter.  The distribution of the bolometric luminosity for the
  comparison sample is approximately normal, with no preference for
  $E1$ (Fig.~\ref{cumulative_global}, while the $E1>0$ ($E1<0$)
  FeLoBALQs have higher  (lower) $L_\mathrm{Bol}$ than average. \label{e1_vs_lbol}} 
\end{center}
\end{figure*}

The distribution of $L_\mathrm{Bol}$ is seen in
(Fig.~\ref{cumulative_global}). It is notable that the
comparison sample $L_\mathrm{Bol}$ distributions split at $E1=0$ are
consistent, but the FeLoBALQ sample $L_\mathrm{Bol}$ distributions for
$E1>0$ and $E1<0$ are significantly different ($p < 0.001$;
Table~\ref{data}).  The median values for the $E1<0$ and $E1>0$
FeLoBALQs are 45.9 and 46.5. The relationship between $E1$ and
  $L_\mathrm{Bol}$ 
is shown in Fig.~\ref{e1_vs_lbol}.  Recalling that the FeLoBAL quasars
were chosen based solely on their spectroscopic classification, this
result seems to be surprising and seems to represent a fundamental
property of low redshift FeLoBAL quasars.

The black hole mass in the $E1>0$ FeLoBAL quasars is lower than in the
$E1<0$ FeLoBAL quasars, but no such significant difference is present
among the comparison objects.   Recall that the black hole mass
estimate depends on luminosity and the H$\beta$ line width so that
more luminous objects with broader lines have larger black-hole
masses.  In our case, the $E1>0$ ($E1<0$) objects have narrower
(broader) H$\beta$ lines and larger (smaller) luminosities so that one
might expect that the $E1>0$ and $E1<0$ objects should have more or
less the same black hole mass distribution.    The 
differences in luminosity and black hole mass are propagated into
$L_\mathrm{Bol}/L_\mathrm{Edd}$, with the $E1<0$ FeLoBAL quasars
accreting at rates more than half a dex lower than the $E1>0$ FeLoBAL
quasars.   This difference seems to be a fundamental one and an
identifying  feature of the two groups of FeLoBAL quasars.

Despite the differences in the black hole mass and Eddington ratio
distributions, the size of the 2800\AA\/ continuum emission region is
the same for the $E1<0$ and $E1>0$ objects.   The emission region size
depends on the temperature distribution as a function of radius.  The
radius for a particular temperature should be larger for a larger
accretion rate, and larger for a larger black hole mass.  The black
hole mass is larger (smaller) for the $E1<0$
($E1>0$) objects, but the accretion rate is smaller (larger), so the
two effects cancel out.

\subsection{Spearman-Rank Correlations}\label{correlation_matrix}
\begin{figure*}[!t]
\epsscale{1.0}
\begin{center}
\includegraphics[width=5.0truein]{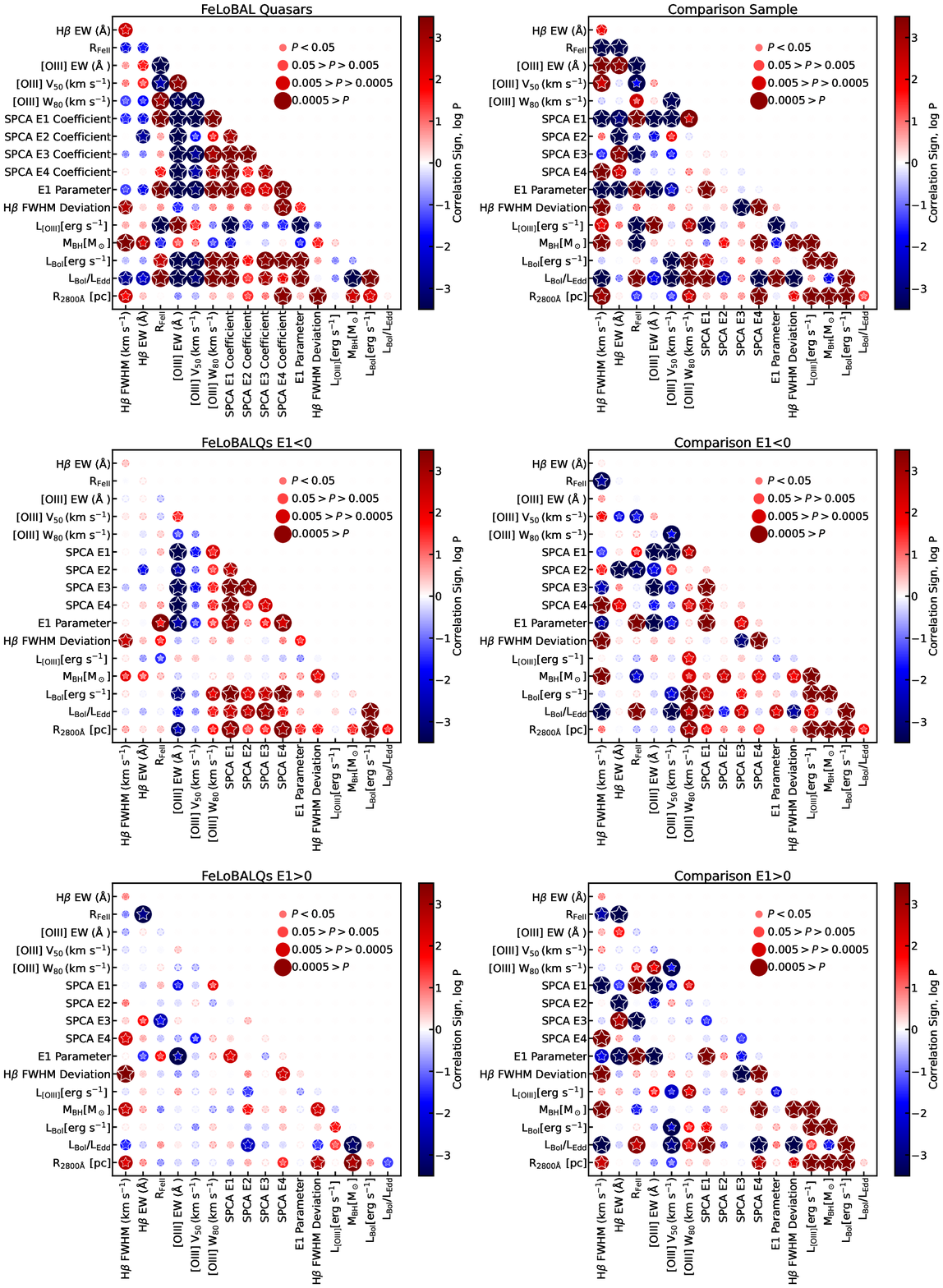}
\caption{The results of the Spearman rank correlation analysis for the
  FeLoBAL quasars (left) and the comparison sample (right).
  The size  and color of the marker indicates the sign and $p$ value
  of the correlation.  Anticorelations are shown in blue, and
  correlations are shown in red.  The saturation of the color 
indicates the significance of the correlation as a continuous
variable, while discrete sizes of the points characterizes a range of
$p$ values as shown in the legend.  The circular markers show the
results for 
  parameter values.  The stars show the results for a Monte Carlo
  scheme to estimate the effects of the errors (see text for
  details). The top panel shows the results for all 30 FeLoBALQs and
  132 comparison sample objects, while the two lower panels show the
  results divided by $E1$ parameter.   Many parameters show similar
  correlations in both panels;   however, some expected correlations,
  for example between the   bolometric luminosity and the [\ion{O}{3}]
  luminosity, are missing   in the FeLoBALQ sample.   \label{correlation}}  
\end{center}
\end{figure*}

The seventeen optical emission line and global properties were
correlated against one another for  both the FeLoBALQ sample and the
comparison sample (Fig.~\ref{correlation}).  The Spearman Rank
correlation is appropriate for these nonparametric data. The plots
represent the log of the $p$ value for the correlation, where the sign
of the value gives the sense of the correlation.  That is, a large
negative value implies a highly significant anticorrelation.

Our method for propagating errors was described in
\S\ref{optical_modeling}.  We chose $p<0.05$ as our threshold for
significance.  The overplotted stars show the fraction of draws that
yield $p$ values greater than our threshold value.  Generally, taking
the errors into account did not affect the significance of a
correlation, if present.  

As discussed in \S\ref{dist_comp},  correlations among quasar
properties associated with the \citet{bg92} Eigenvector 1, here
parameterized using the $E1$ parameter, dominate and potentially
obscure other correlations.  We would like to determine whether there
were correlations independent of the $E1$ parameter.  It is also
possible that the two classes of FeLoBALQs show different correlation
behaviors  that would be washed out in the 
whole-sample correlations.  Therefore, we also compute the
Spearman-rank correlation coefficient between the parameters divided
by $E1<0$ and $E1>0$ (Fig.~\ref{correlation}, middle and lower
panels). 

Some of the correlations we observe are a consequence of parameter
dependence. Other correlations are very well known; for example the 
anticorrelation between $R_\mathrm{FeII}$ and the [\ion{O}{3}]
equivalent width. The origin of these correlations has been discussed
exhaustively in the literature \citep[e.g.,][and references
therein]{shen_ho_14}.  Notable correlations and other patterns in the
data are discussed in the sections below.

{ As discussed in \S\ref{optical_modeling}, the [\ion{O}{3}] and
\ion{Fe}{2} emission is weak in some objects, and it is possible to
consider the measurements of those variables that are not
found to be significant using the F test to be
upper limits.  The generalized Kendall Tau test can be used for
censored data \citep{isobe86}.  We use the {\tt pymccorrelation}
implementation \citep{privon20} on the [\ion{O}{3}] equivalent width, 
R$_\mathrm{FeII}$ and [\ion{O}{3}] luminosity, and looked for
parameter combinations that either switched from being a significant
correlation ($P<0.05$) to insignificant or vice versa.  For the
FeLoBALQs (including all, $E1>0$, and $E1<0$ combinations), 13
correlations (9\% of the 144 non-trivial correlations for these three
parameters) changed, with 4 becoming insignificant and 9 becoming
significant.  For the  comparison sample (including all, $E1>0$, and
$E1<0$), 12 correlations (8\%) changed, with 4 becoming insignificant
and 8 becoming significant.  We credit these minor changes to the fact
that the upper limits only strengthen the observed anticorrelation
between [\ion{O}{3}] and R$_\mathrm{FeII}$ that defined $E1$ because
[\ion{O}{3}] (\ion{Fe}{2}) was found to statistically unnecessary only
among the $E1>0$ ($E1<0$) objects.} 

\subsubsection{Shen \& Ho 2014}\label{shen_ho_section}

\citet{shen_ho_14} presented an  analysis of $\sim20000$ low-redshift
quasars from the SDSS DR7 quasar catalog.  They asserted 
that $R_\mathrm{FeII}$ is a measure of the accretion rate relative to
the Eddington value (the Eddington ratio), with large values of this
parameter corresponding to a large Eddington ratio.  They found a wide
range of values of H$\beta$ FWHM for a given value of
$R_\mathrm{FeII}$ and suggested that the dispersion of the H$\beta$
FWHM for a given value of $R_\mathrm{FeII}$ reflects a range of
viewing angle orientations to a disk-like broad-line region.   

\begin{figure*}[!t]
\epsscale{1.0}
\begin{center}
\includegraphics[width=6.5truein]{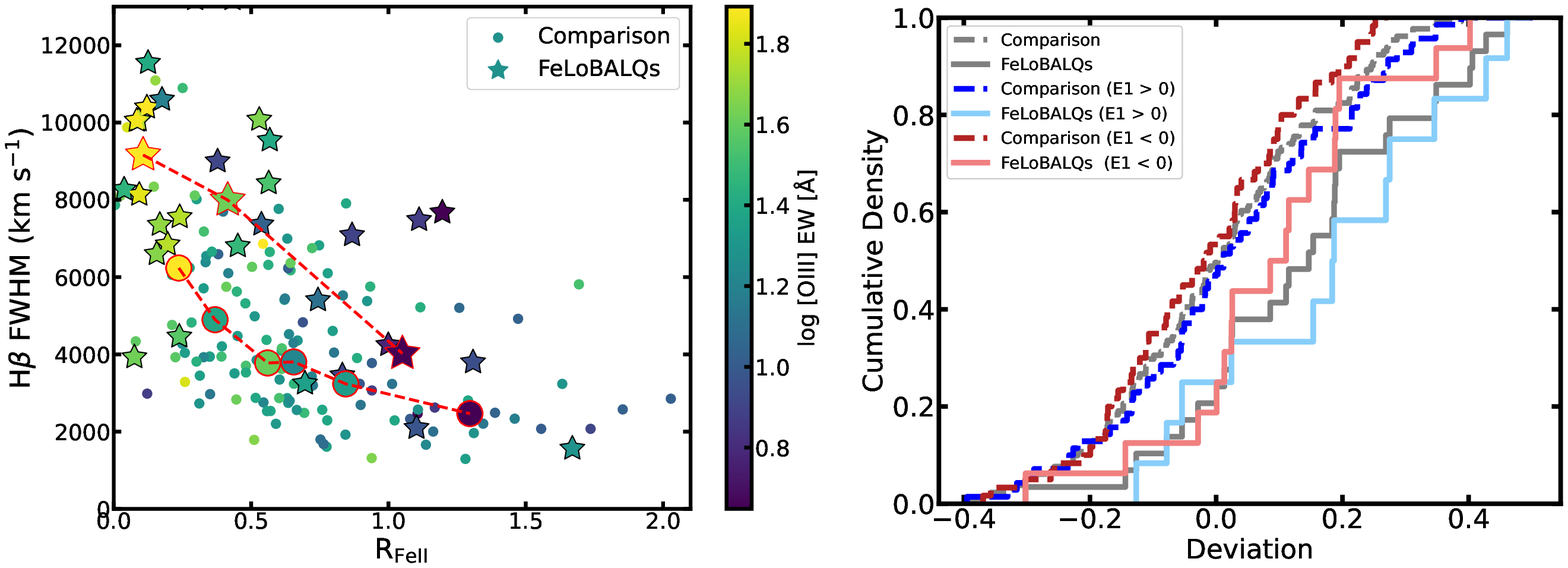}
\caption{{\it Left:} The H$\beta$ FWHM as a function of
  $R_\mathrm{FeII}$ for the 
  FeLoBALQs and comparison sample following \citet{shen_ho_14}.  The
  large points outlined in red   show the median H$\beta$ FWHM values
  in bins of $R_\mathrm{FeII}$. 
  It is clear that the H$\beta$ FWHM is systematically broader in the 
  FeLoBALQs for a given value of $R_\mathrm{FeII}$. {\it Right:} The
  cumulative distributions of the deviation parameter. The color
  scheme is the same as used Fig.~\ref{plot_comp_1}. The deviation
  parameter separates the FeLoBALQs from the unabsorbed comparison
  sample more cleanly than does the H$\beta$ FWHM
  (Fig.~\ref{plot_comp_1}). \label{shen_ho_fig}}     
\end{center}
\end{figure*}

We reproduced their Figure 1 for the FeLoBALQs and comparison sample
in the left panel of Fig.~\ref{shen_ho_fig}.  We already know from
\S\ref{optical} and 
Fig.~\ref{plot_comp_1} that the H$\beta$ FWHM is systematically
broader among the FeLoBALQs.  This representation shows that the
difference is a function of $R_\mathrm{FeII}$ such that objects with
larger (smaller) values of $R_\mathrm{FeII}$ have systematically
smaller (larger) H$\beta$ FWHM.  We highlight the difference by
computing the median value in bins of $R_\mathrm{FeII}$ such that each
bin has the same number of points: six bins of 22 objects each for the 
comparison sample and three bins of 10 objects each for the FeLoBALQs.   

To quantify the difference, we defined a parameter called H$\beta$
FWHM deviation, abbreviated hereafter as deviation.   To compute the
deviation for each point, we  first linearly interpolate the medians
computed above for the comparison sample.  For points with larger
$R_\mathrm{FeII}$ values than 
the the median of the largest or smallest bin, we used the two nearest 
median points to extrapolate.  The deviation was then defined to be
the difference between the log of the observed H$\beta$ FWHM and the
log of the interpolated relationship.  \citet{shen_ho_14} found a
lognormal distribution of H$\beta$ FWHM as a function of
$R_\mathrm{FeII}$ with a dispersion of 0.15--0.25 dex. The disperson
in the deviation for the comparison sample is 0.18 and is thus
consistent with the \citet{shen_ho_14} value.  

The cumulative distribution of the deviation is shown in
Fig.~\ref{shen_ho_fig}.  It is useful to compare this plot with the
H$\beta$ FWHM cumulative distribution (Fig.~\ref{plot_comp_1}).  The
median values of FWHM for the $E1<0$ and $E1>0$ groups are dramatically
different.  This result is now explained 
since the two groups are characterized by much different values of
$R_\mathrm{FeII}$, since $R_\mathrm{FeII}$ contributes to the
definition of the $E1$ parameter.   The deviation parameter takes this
difference into account since it is function of $R_\mathrm{FeII}$.
The cumulative distributions show that median deviation for the
comparison sample is consistent with zero (by definition), and the
median value for the FeLoBALQs is 0.17, corresponding to 47\%
systematically larger H$\beta$ FWHM when averaged over all the
objects.    

\citet{shen_ho_14} conjecture that the deviation in H$\beta$ FWHM at
each value of $R_\mathrm{FeII}$ is a consequence of differences in
orientation.  If true, our result implies the FeLoBALQs are observed
at systematically larger inclination angle than objects without
absorption.  If we take the effective mean inclination with respect to
the normal of unabsorbed
objects to be 30 degrees \citep[e.g.,][]{shen_ho_14}, then assuming
the same geometry for the broad-line region and a $\sin \theta$
dependence, we find that the effective mean inclination to the
broad-line quasars should be 48 degrees.  The FWHM enters the virial
black hole mass as $FWHM^2$, so a 47\% larger value suggests that the
black hole masses for the FeLoBALQs are over-estimated by a factor of 2.2  
on average. 

A larger inclination viewing angle might be expected for BALQs.  Due
to the requirement that angular momentum be dispersed for accretion to 
occur, the accreting gas is expected to be disk-like, not only at the 
smallest scales (i.e., the optical-UV emitting accretion disk) but
also at larger scales (the molecular torus and beyond).  If that
material is transported perpendicular to the disk by some local
mechanism (i.e., radiative line driving, scattering on dust), then
once the material rises sufficiently to be illuminated by the
brilliant quasar nucleus, the stream lines should be bent radially so
that the observer views the quasar nucleus through absorbing gas only
along certain lines of sight \citep[e.g.,][]{elvis00}.   

Although promising, a larger inclination viewing angle falls short
of explaining all of the differences between the FeLoBALQs and the
comparison sample.  For example, it does not explain the FeLoBALQ
$E1$ bimodality.  Moreover, the explanation is not unique.
A broader H$\beta$ emission line would also be produced if the
broad-line region were truncated at its outer boundary.  Quasars with
double-peaked broad lines are extreme examples of such objects
\citep[e.g.,][]{strateva03}, and \ion{Fe}{2} absorption has been found
in several \citep{halpern96, halpern97, eracleous03}. We favor this
interpretation and return to this idea in Paper IV (Leighly et al.\ in
prep.).  

\subsubsection{Boroson 2002}\label{boroson02}

\citet{bg92} presented a principal components analysis of measured
emission-line and continuum properties near H$\beta$ in a sample of
87 low-redshift quasars from the Bright Quasar Survey \citep{hewett95}.
\citet{boroson02} used these results augmented by results from a
sample of radio-loud objects to investigate the role of the PCA
eigenvectors in determining an object's class, i.e., whether an object
is radio loud or radio quiet.   His sample included a small number of
broad absorption  line quasars.    His discussion culminates in his
Fig.\ 7, an  interpretive diagram that used principal components 1 and
2 to classify AGN and quasars.  

\begin{figure*}[!t]
\epsscale{1.0}
\begin{center}
\includegraphics[width=6.5truein]{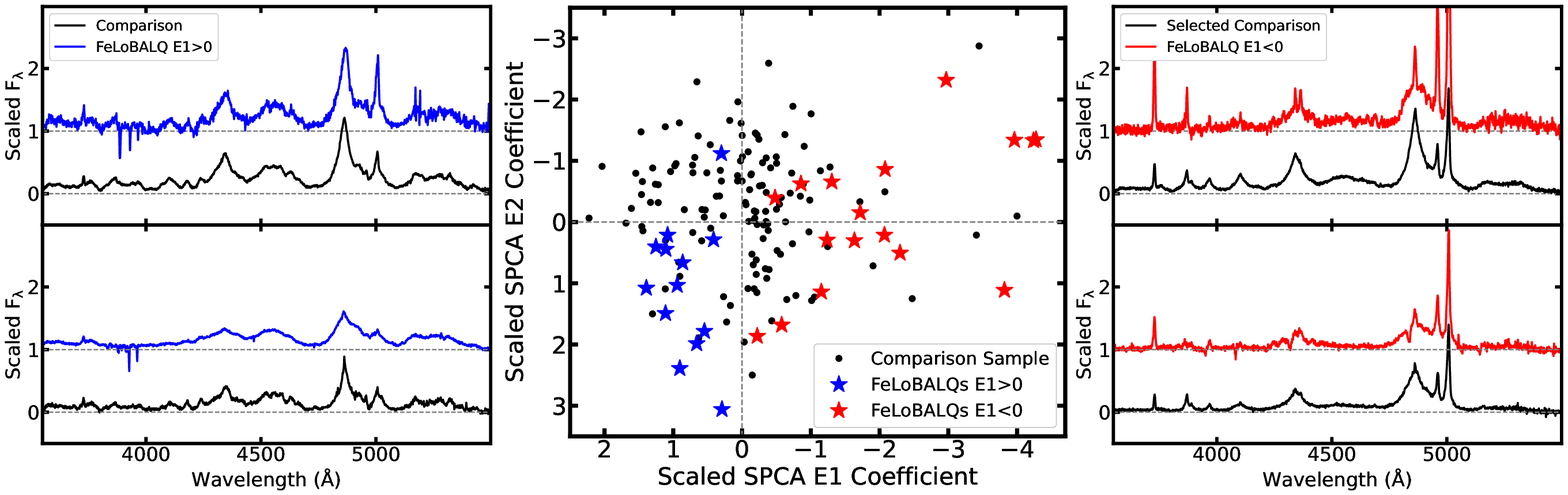}
\caption{Plots modeled after Fig.~1 and Fig.~2 in \citet{boroson02}.
  The central panel shows the spectral PCA fit coefficients scaled by
  the variance of the comparison sample in each quadrant. The right
  and left   panels show weighted-mean composite spectra from each quadrant
  offset in the $y$ direction for plot clarity. The
  highest concentration of FeLoBALQs relative to comparison objects is
  found in the lower left quadrant, as predicted by \cite{boroson02};
  however, all of the $E1<0$ objects lie in the two right
  quadrants where \cite{boroson02} found radio-loud quasars. 
   \label{boroson_first_fig}}       
\end{center}
\end{figure*}

We present a similar analysis to determine whether the same pattern of 
behavior is present among our FeLoBALQs and the comparison sample.  
The difference is that we used the coefficients derived from model
fitting using the spectral principal components
analysis eigenvectors created from the comparison sample
(\S\ref{pca}).  PCA eigenvector coefficients are invariant 
under sign change, so we use the composite spectra shown in
\citet{boroson02} Fig.\ 2 to orient ours.  The results are shown in
Fig.~\ref{boroson_first_fig}.  The negative of our SPCA1 coefficients
correspond well with the \citet{boroson02} PC1 coefficients.  The
SPCA2 coefficients correspond less well with the \citet{boroson02} PC2
coefficients, so we used the [\ion{O}{3}] equivalent width in the
composite spectra to assign the negative of our SPCA2 coefficients
to the \citet{boroson02} PC2 coefficents.
Fig.~\ref{boroson_first_fig} (center) shows the SPCA2 eigenvector 
coefficients as a function of the SPCA1 eigenvector coefficients.  To
make the plot more similar to Fig.~1 in \citet{boroson02}, the SPCA
coefficients were scaled by dividing by the the standard deviation of the comparison sample in each quadrant.  The FeLoBALQ coefficients 
are also plotted; their SPCA values have been scaled with the
same factors as the comparison sample.  

This figure highlights some interesting differences between the
comparison objects and the FeLoBAL quasars.  There is a  striking
difference between comparison and FeLoBALQ number density 
in the left two quadrants.  The upper left quadrant includes 30\% of the
comparison objects, but only one FeLoBALQ, while the lower left
quadrant includes 40\% of the FeLoBALQs, but only 11\% of the
comparison objects.  Thus there is a strong difference in SPCA2
between the comparison objects and the FeLoBALQs.  Examining the
composite spectra in the left two panels in
Fig.~\ref{boroson_first_fig} and the PCA eigenvectors in
Fig.~\ref{pca_eigen} shows that the difference is that the FeLoBALQs
have weaker and broader Balmer emission lines, a property that should
correlate with SPCA2.  As noted above, a physical interpretation of
  this result is not obvious.  

\citet{boroson02} used his Fig.~1 and ensuing discussion to construct
a schematic interpretive diagram based on PC1 and PC2 (his Fig.~7).
He argued that PC1 is a measure of the Eddington ratio, and PC2 is
a measure of the accretion rate which is equivalent to the luminosity
for a constant accretion efficiency.  Lines of
constant black hole mass lie along the diagonal.   He found
that a plot of PC2 versus  PC1 distinguishes between radio-loud and
radio-quiet quasars.  He concluded that narrow-line Seyfert 1 galaxies
and broad absorption line quasars (both generally radio quiet) are
characterized by high values of the Eddington ratio, with the BAL
quasars additionally characterized by a high accretion rate.
Radio-loud objects have larger black hole masses and lower accretion
rates. 

Comparing the central panel of Fig.~\ref{boroson_first_fig} with
\citet{boroson02} Fig.~7 shows that the density of BAL
quasars in our sample is high in the lower left quadrant as
predicted.  A difference is that all of the $E1<0$ FeLoBALQs lie on the  
right side of the diagram, as they should, since $E1$ is strongly
correlated with the SPCA1 coefficient.  The right side of the diagram
is where the \citet{boroson02} radio-loud objects are. Some of our
FeLoBALQs may be radio loud; a  systematic investigation of the radio
properties of this sample is beyond the scope of this paper.  However,
we note that SDSS~J102226.70+354234.8 has a resolved two-sided jet and
$E1=0.97$, and SDSS~J164419.75+530750.4 has a 500~kpc scale radio lobe with
bilateral hotspots and $E1=-0.51$ (L.\ Morabito, P.\ Comm., 2021).    

\begin{figure*}[!t]
\epsscale{1.0}
\begin{center}
\includegraphics[width=4.5truein]{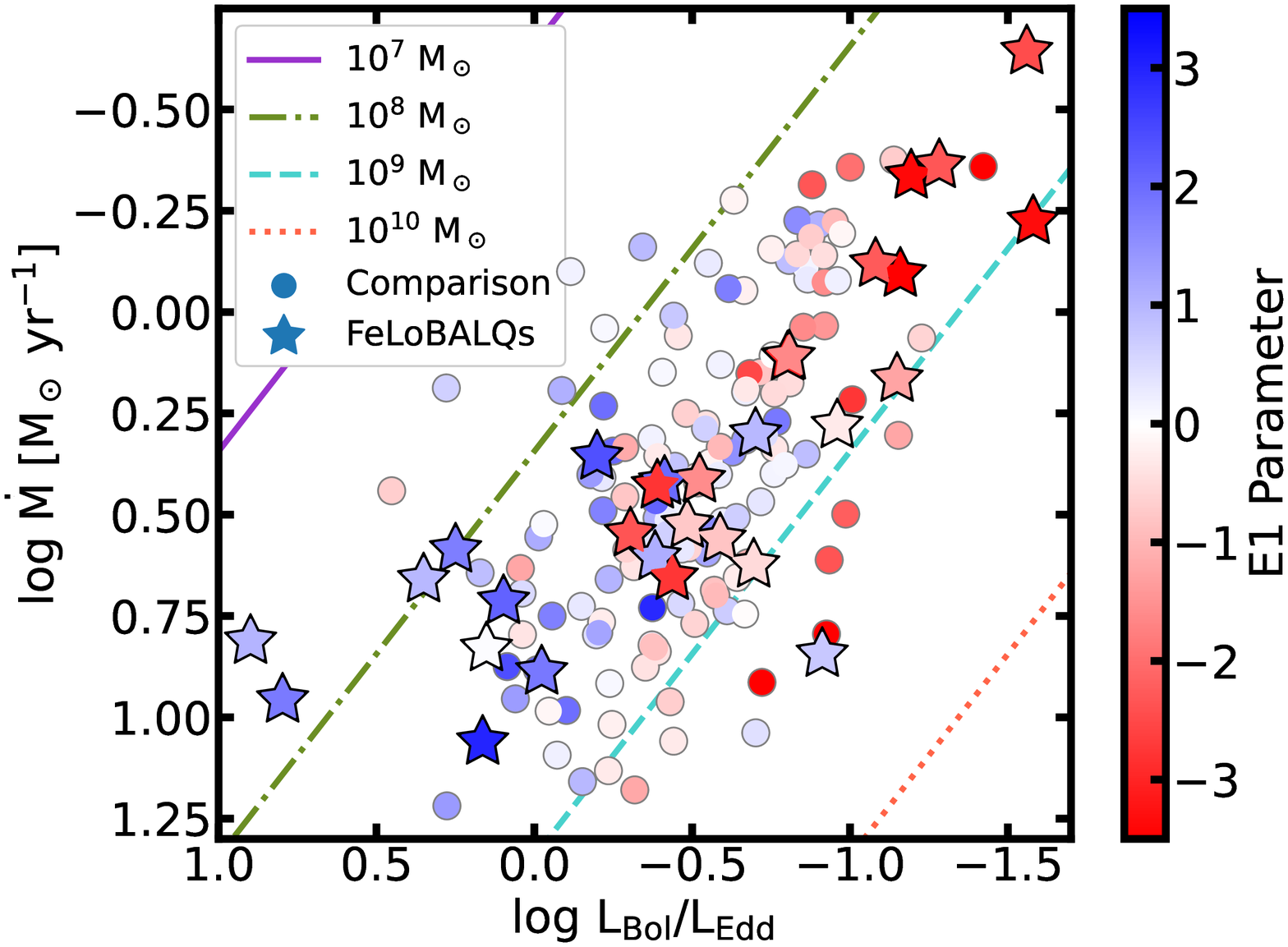}
\caption{A plot modeled after Fig.~7 in \citet{boroson02} rendered
  using our estimates  of  the Eddington ratio and accretion rate.
  \citet{boroson02} predicted that BALQs should lie in the lower left
  corner of this plot, i.e., high Eddington ratio.   We found that the
  FeLoBALQs are characterized by both low and high Eddington ratios,
  although their distribution in this parameter is statistically
  different from the distribution of the comparison objects; see
  Fig.~\ref{hist_ave_spec}, Fig.~\ref{plot_comp_3},
  Table~\ref{distrib_tab}.      \label{boroson_lum_mdot}}     
\end{center}
\end{figure*}

Because we have measured the Eddington ratio and can compute the
accretion rate, we can directly investigate whether our principal
components coefficients correlate with these properties and render
the \citet{boroson02} Fig.\ 7 using our measurements
(Fig.~\ref{boroson_lum_mdot}).   As shown in Fig.~\ref{correlation},
SPCA1 is strongly positively correlated with the Eddington ratio
(comparison: $p=6.4 \times 10^{-10}$; FeLoBALQs: $p=3.2 \times
10^{-8}$).  The accretion rate was computed using 
$L_\mathrm{Bol}=\eta \dot{M} c^2$, where we assumed a constant
$\eta=0.1$, acknowledging that this value may not be accurate at very
low or very high accretion rates.  The accretion rate is strongly
correlated with the SPCA1 coefficients (comparison: $p=0.0022$;
FeLoBALQ: $p=1.1 \times 10^{-6}$).  It is not strongly correlated with
SPCA2 (comparison: $p=0.6$; FeLoBALQ: $p=0.014$).  Thus the x-axis in
our figure and the \citet{boroson02} Fig.\ 7 both correspond to the
first eigenvector; this is further illustrated by the color bar
corresponding to $E1$ parameter value in Fig.~\ref{boroson_lum_mdot}.   
However, our  y-axis does not correspond with the second
eigenvector.  So although we assigned SPCA2 to the $y$ axis in
Fig.~\ref{boroson_first_fig} based on the features in the composite
spectra, we did not find that SPCA2 is a measure of the accretion rate,
although it does very well separate the $E1>0$ FeLoBALQs from the
comparison objects.   

In addition, Fig.~\ref{boroson_lum_mdot} shows that in
our sample the FeLoBAL quasars and comparison sample objects are
distributed roughly uniformly in Eddington ratio versus accretion rate
space (although statistically distinguishable; see
Fig.~\ref{hist_ave_spec}, Fig.~\ref{plot_comp_3},
Table~\ref{distrib_tab}), and are not  relegated to high Eddington
ratios and high accretion rate, as proposed by \citet{boroson02}. 

We note that other recent work in the UV rest wavelengths does not
support the  
classification scheme proposed by \citet{boroson02}.  There is
evidence that the blueshift and equivalent width of the \ion{C}{4}
emission line is associated with the \citet{bg92} eigenvector 1
\citep[e.g.,][]{wills99,shang03}, and therefore by extension the
Eddington ratio.  \citet{rankin20} found that BAL quasars exist
throughout \ion{C}{4} blueshift / equivalent width space, and
\citet{rankine21} showed that radio-loud quasars similarly populate
this space.  We used \citet{boroson02} for our discussion because he
used the rest-frame optical eigenvectors and is therefore more
similar to our analysis.

\subsubsection{Correlations Among Global
  Properties}\label{global_corr}

Next we examined correlations among the global parameters: the black
hole mass, bolometric luminosity, and accretion rate relative to
Eddington.

We found that the E1 parameter is strongly positively correlated with
$L_\mathrm{Bol}$ among the FeLoBAL quasars ($p=4.4\times 10^{-5}$),
but no such correlation is present in the comparison sample
($p=0.15$).  This result echos Fig.~\ref{e1_vs_lbol}, which shows that
the $E1<0$ FeLoBALs are less luminous than the $E1>0$ FeLoBALs
(\S~\ref{twopop}).  Thus, the optical spectral properties of FeLoBAL
quasars in this sample predict the luminosity.  This result is
surprising and unexpected. 

\begin{figure*}[!t]
\epsscale{1.0}
\begin{center}
\includegraphics[width=6.5truein]{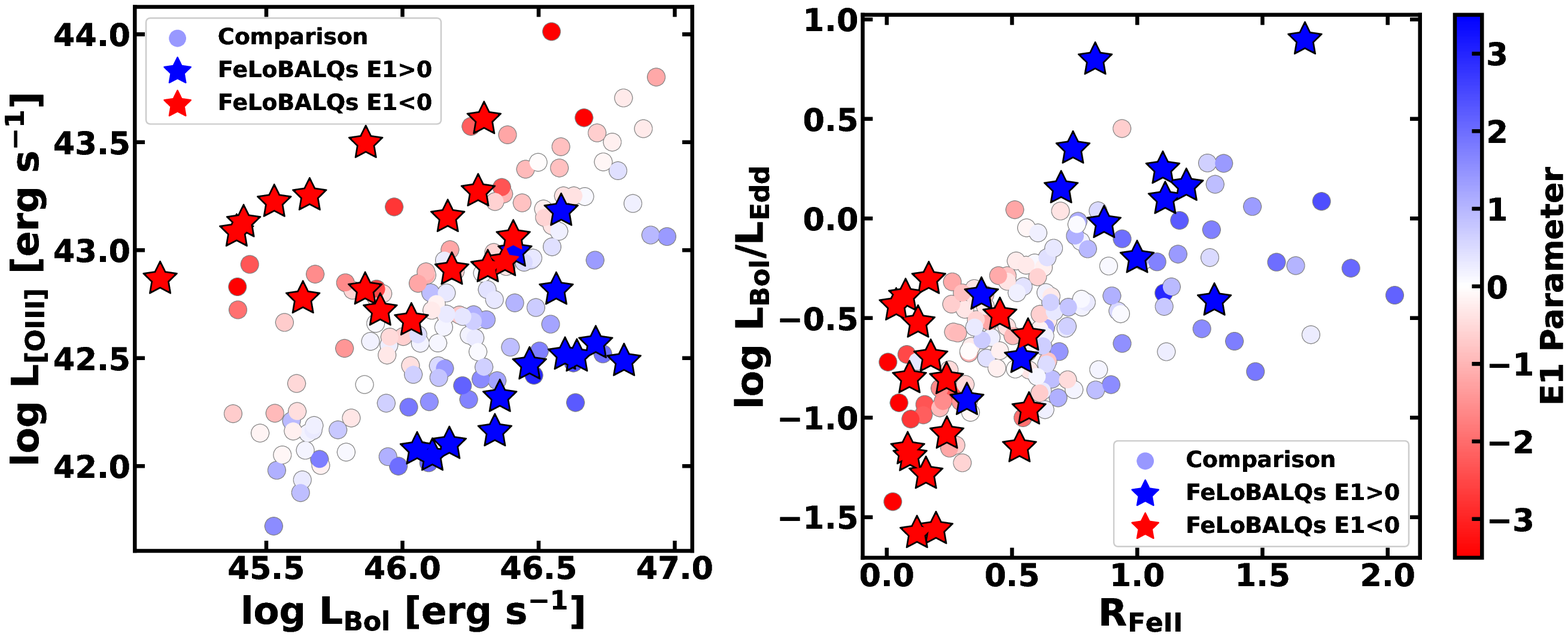}
\caption{Relationships among global parameters.  {\it Left:}  A strong
  positive correlation is observed between the bolometric luminosity
  and the [\ion{O}{3}] luminosity for the comparison sample
  ($p=5.2\times 10^{-19}$), but not
  for the FeLoBALQs ($p=0.12$).  Rather, they form two rough tracks of
  approximate correlation on either side of the comparison sample
  objects and divided by $E1$ parameter.  {\it Right:} The computed
  ratio of the bolometric  luminosity to the Eddington luminosity as a
  function of   R$_\mathrm{FeII}$.  The two parameters are
  correlated ($p=1.2\times 10^{-4}$ for the FeLoBALQs, and
  $p=7.1\times 10^{-11}$ for the comparison sample).      \label{global}}      
\end{center}
\end{figure*}

We also investigate the [\ion{O}{3}] luminosity.  
There is a correlation between the bolometric luminosity and the
[\ion{O}{3}] luminosity in the comparison sample ($p=5.2\times
10^{-19}$).  This type of 
correlation is expected in a flux-limited sample.  The interesting
fact is that no correlation is found between these parameters for the
FeLoBALQs ($p=0.12$).  Fig.~\ref{global} shows the reason for this
behavior. While the correlation in the comparison sample is clearly
seen, the FeLoBALQs instead form two (or more) clumps on either side
of the comparison sample regression.   This behavior further
underlines the presence of two types of FeLoBALQs in this sample.

The R$_\mathrm{FeII}$ parameter can be taken as a proxy for the accretion
rate relative to Eddington \citep[e.g.,][]{shen_ho_14}.
Fig.~\ref{global} shows the relationship between these two
parameters.  A correlation exists, but it is rough although
statistically significant for both the FeLoBALQs and the comparison
sample ($p=1.2\times 10^{-4}$ and $p=7.1\times 10^{-11}$,
respectively).  Why are these two parameters not perfectly correlated?  
 Several approximations enter into the computation of 
$L_\mathrm{Bol}/L_\mathrm{Edd}$.  $L_\mathrm{Edd}$ depends on the
black hole mass.  The 
black hole mass was estimated using the FWHM of the H$\beta$ line and
following the formalism of \citet{collin06}.  In particular, we used
their Eq.\ 7 to estimate the  scale factor $f$ based on the FWHM of
the H$\beta$ line.  This assumption must be a simplification given the
systematically broader H$\beta$ lines observed among the FeLoBAL
quasars, whether that property originates in a more inclined viewing
angle or a lack of a traditional BLR component (or both).  If the
origin of the broader H$\beta$ lines is viewing angle inclination, then the
simplification  would bias the black hole masses larger than they
should be.   Also, we used the measured $L_{5100}$
luminosity to compute the radius of the broad-line region without
correcting for reddening.  A too-small $R_\mathrm{BLR}$ would bias the
black hole mass smaller than they should be.  

The bolometric luminosity was computed using a bolometric correction
factor from the 3$\mu \mathrm{m}$ luminosity density
\citep{gallagher07}. That bolometric correction factor was computed
from relatively luminous quasars which typically show near-infrared
emission from a torus. Thus, objects  that are weak at three microns
will potentially have underestimated bolometric luminosities.  In Paper IV,
we investigated the optical to IR spectral energy distribution and
found that some $E1<0$ objects are weak at 3 microns.  

The combination of uncertainties in the black hole mass and bolometric
luminosity estimates must contribute to the scatter in
$L_\mathrm{Bol}/L_\mathrm{Edd}$ seen in the right panel of
Fig.~\ref{global}.  It is interesting that among the $E1<0$ FeLoBALQs,
$L_\mathrm{Bol}/L_\mathrm{Edd}$ is correlated with $L_\mathrm{Bol}$
but not $M_\mathrm{BH}$ ($p=5.6\times 10^-6$ and $p=0.36$,
respectively) , while the opposite is true for the $E1>0$
FeLoBALQs ($p=0.18$ and $p=1.2 \times 10^{-4}$, respectively;
Fig.~\ref{correlation}).  It is possible that, in this sample,
R$_\mathrm{FeII}$ better reflects the physical conditions of 
the nucleus anticipated to depend on accretion rate than does the
computed value of $L_\mathrm{Bol}/L_\mathrm{Edd}$.  It would be
interesting to obtain stellar velocity dispersion black hole estimates
to substantiate this inference.  

A related question is, how interchangable are $E1$, the first SPCA
coefficient, $R_\mathrm{FeII}$ and the Eddington ratio?  All of these
parameters are highly correlated (Fig.~\ref{correlation}); the largest
$p$ value among all six combinations is $1.2\times 10^{-4}$ ($9.3\times
10^{-7}$) for  the FeLoBALs (comparison sample).  Yet the $E1$
parameter is the only one of the four that passes the GMM test for 
bimodality (\S~\ref{twopop}), i.e., a distribution with negative
kurtosis and a better fit with two heteroskedastic Gaussian
distributions than with one, even though bimodality is convincingly
established since the distribution fails the non-parametric dip test.
At least $R_\mathrm{FeII}$ and the Eddington ratio have distributions
with negative kurtosis.     Perhaps this implies that the $E1$
parameter captures an essence of the data that is blurred among the
other parameters.

\section{Discussion and Conclusions}\label{discussion}

We presented analysis of the rest-optical spectra near H$\beta$ of a
unique sample of low redshift ($z<1$) FeLoBAL quasars.  All objects
have sufficient quality data that the H$\beta$ / [\ion{O}{3}] region of the
spectrum could be analyzed.  Many of the  objects were selected using
a convolutional neural net (Dabbieri et al.\ in prep.) applied to
$0.8<z<1$ quasars from the DR14 quasar catalog
\citep{paris18}, and of the thirty objects, { eleven} had not been
classified as a BAL quasar previously.  A 132-object comparison
sample, matched in redshift, median signal-to-noise ratio in the
H$\beta$ region, and 3-micron luminosity density was compiled and
analyzed in parallel. 

BAL quasars can be difficult to find and recognize compared with
unabsorbed quasars, and therefore construction of a uniform sample is
difficult.  The SDSS quasar catalogs require the presence
of a broad emission line in the spectrum, and therefore can be biased 
against heavily absorbed quasars; for example, the spectacular FeLoBAL
quasar SDSS~J135246.37$+$423923.5 \citep{choi20} is not in the DR14
quasar catalog.  However, our selection process reduces some of the
pitfalls of other samples. Although the sample is not large, it is
larger than any samples of rest-frame H$\beta$-region spectra of BAL
quasars considered to date \citep[][16, 8, 16, respectively]{yw03,
  runnoe13, schulze17}.  The sample 
may be more homogeneous, since we chose only FeLoBALQs.  BAL quasar
properties such as outflow velocity show strong luminosity dependence
\citep[e.g.,][]{laor02,ganguly07}. Because our sample spans only a
small range of redshift, the range of luminosities is small. The SDSS
spectra have more uniform quality than near-infrared
spectra can be, due to variable-quality telluric correction and
other factors such as fixed pattern noise that degrade the spectra.
The ability to compile a large and uniform comparison sample that can
be analyzed uniformly turned out to be critical for many of the
conclusions in the paper. 

Our analysis focused on measurement of the H$\beta$, [\ion{O}{3}], and
\ion{Fe}{2}.  We used traditional multi-component model fitting, as
well as a principal components analysis.  We found that the variance
in the emission lines in the comparison sample was principally
associated with the \citet{bg92} Eigenvector 1, as expected.  We
defined an empirical parameter $E1$ which is a function of the
[\ion{O}{3}] equivalent width and $R_\mathrm{FeII}$, the ratio of the
\ion{Fe}{2} emission to the H$\beta$ emission
\citep[e.g.,][]{shen_ho_14}.  Both $E1$ and $R_\mathrm{FeII}$ were
found to be strongly correlated with the Eddington ratio
(Fig.~\ref{correlation}, \ref{global}).  A large and positive $E1$
parameter corresponds to a high Eddington ratio, while a large an
negative $E1$ parameter corresponds to a low Eddington ratio.  

The large and well-defined matched sample of unabsorbed quasars
allowed us to test whether low-redshift FeLoBAL quasars have the same
rest-frame optical emission line properties as unabsorbed objects.
The principal result of the paper is that they do not.   The
uanbsorbed quasars have a peaked distribution in the
$E1$ parameter (Fig.~\ref{hist_ave_spec}). In contrast, the FeLOBALQs
show an apparent bimodal distribution in $E1$, although not
statistically significant.  Moreover, the luminosities of the
two groups of FeLOBALQs is different (Fig.~\ref{cumulative_global},
Fig.~\ref{e1_vs_lbol}), and those differences carry over to a
difference in Eddington ratio (Fig.~\ref{cumulative_global}).  
The implication is that the low-redshift FeLoBALQs are characterized
by either a large Eddington ratio, or a small one, but not an
intermediate value.  This division is echoed in the relationship
between the [\ion{O}{3}] and $L_\mathrm{Bol}$.  These parameters are
correlated in the comparison sample, but clustered by $E1$ in the
FeLoBAL quasars (\S\ref{global_corr}; Fig.~\ref{global}).   

This result is new and distinct from what has been previously reported
in the literature.  Less work has been done in the rest-frame optical
band compared with the rest-frame UV; nevertheless, it has been reported that
BAL quasars appear to be either high Eddington ratio objects
\citep{yw03,runnoe13}, or that there is no difference between absorbed
and unabsorbed objects \citep{schulze17}.  The difference between our
results and the previous ones likely stems from the large and uniform
parent sample: the low redshift SDSS quasar catalog, supplemented by
FeLoNET, our convolutional neural net, used to discover more than half
of the FeLoBALQs analyzed here. 
This result inspires the question: why does FeLoBAL absorption occur
in objects with high and low accretion rates, but not in objects with
an intermediate accretion rate?  This question is addressed in
Paper IV (Leighly et al.\ in prep.).

We found that the measurements of H$\beta$ FWHM of the FeLoBALQs are
consistently larger than those of the comparison sample
(Fig.~\ref{plot_comp_1}).  Moreover, they are consistently larger for
the same value of $R_\mathrm{FeII}$ (\S\ref{shen_ho_section},
Fig.~\ref{shen_ho_fig}). \citet{shen_ho_14} proposed that
$R_\mathrm{FeII}$ correponds to Eddington ratio, and that the scatter
in H$\beta$ width indicates objects with different inclination angles.
If this interpretation is correct for the FeLoBAL quasars, then if
normal quasars are typically observed at 30 degrees from normal to the
accretion disk, the FeLoBALQs are typically observed at 48 degrees.
However, this reasoning does not explain all of the many differences
we found between the FeLoBAL quasars and the unabsorbed comparison
sample; for example, it  does not explain the apparent bimodality in
the FeLoBAL distribution.  Moreover, this
explanation is not unique.  Instead, the narrow core of
the H$\beta$ emission line may be missing.  Additional support for
this idea is presented and discussed in Paper IV (Leighly et al.\ in
prep.). 

We compared our results with those of \citet{boroson02}.   He found
evidence that the rest-frame optical band PCA eigenvector coefficients could be
used to separate quasars among narrow-line Seyfert 1 galaxies, BAL
quasars and radio-loud objects.    In particular, he
found that BAL quasars were relegated to a corner of the PCA1-PCA2 coefficient
graph.  We confirmed that nearly all of our $E1>0$ (high Eddington
ratio) FeLoBAL quasars fell in that region of parameter
space. However, we found that all of our $E1<0$ were scattered among
the radio-loud objects in the \citet{boroson02} scheme.  We are in the
process of investigating the radio properties of this sample.   We
also note that more recent work using rest-frame UV spectra has found
a more uniform distribution of types of objects
\citep[e.g.,][]{rankin20,   rankine21}. 

Moving forward, we note that the \ion{Fe}{2} absorption spectrum
blankets the rest-frame UV region 
and FeLoBAL quasars can be observed to resdshifts higher than $z\sim
3$.  In a flux-limited sample like the SDSS, higher redshift samples
are comprised of higher luminosity objects.  Assuming fixed Eddington 
ratio distribution, such objects will have larger black hole masses.
A larger black hole mass will have a softer (UV dominant) spectral
energy distribution, which could affect the outflow in two ways.  The
softer SED might produce an outflow with a different ionization
balance, i.e., a larger fraction of lower-ionization species, which
means the observed absorption could change.  A dramatic change is
probably not expected; the observed absorption lines should depend
principally on the ionization parameter, as is usually the case.  The
softer SED should be more efficient in accelerating  outflows, since
it has more photons with wavelengths suitable for resonance scattering
and fewer likely to over-ionize the outflowing gas.  In fact, we see
some evidence for faster outflows and shifted parameters among the
higher redshift FeLoBAL quasars we have analyzed already
\citep{voelker21}. We also expect that lower-Eddington-ratio objects
will be rarer \citep[e.g.,][]{jester05}.      This is because the
black hole mass function is steep and objects with black holes larger
than $10^{10}\rm \, M_\odot$ are very rare.

\acknowledgements

KML acknowledges very useful conversations with Leah Morabito, and
thanks Alex Parsells for running the \citet{hartigan85} dip test.
Support for {\it SimBAL} development and analysis is provided by NSF
Astronomy and Astrophysics Grants No.\ 1518382 and 2006771. This work
was performed in part at Aspen Center for Physics, which is supported
by National Science Foundation grant PHY-1607611. SCG thanks the
Natural Science and Engineering Research Council of Canada.

Long before the University of Oklahoma was established, the land on
which the University now resides was the traditional home of the
“Hasinais” Caddo Nation and “Kirikiris” Wichita \& Affiliated
Tribes. This land was also once part of the Muscogee Creek and
Seminole nations.

We acknowledge this territory once also served as a hunting ground,
trade exchange point, and migration route for the Apache, Comanche,
Kiowa and Osage nations. Today, 39 federally-recognized Tribal nations
dwell in what is now the State of Oklahoma as a result of settler
colonial policies designed to assimilate Indigenous peoples.

The University of Oklahoma recognizes the historical connection our
university has with its Indigenous community. We acknowledge, honor
and respect the diverse Indigenous peoples connected to this land. We
fully recognize, support and advocate for the sovereign rights of all
of Oklahoma’s 39 tribal nations.

This acknowledgement is aligned with our university’s core value of
creating a diverse and inclusive community. It is our institutional
responsibility to recognize and acknowledge the people, culture and
history that make up our entire OU Community.

Funding for the SDSS and SDSS-II has been provided by the Alfred
P. Sloan Foundation, the Participating Institutions, the National
Science Foundation, the U.S. Department of Energy, the National
Aeronautics and Space Administration, the Japanese Monbukagakusho, the
Max Planck Society, and the Higher Education Funding Council for
England. The SDSS Web Site is http://www.sdss.org/.

The SDSS is managed by the Astrophysical Research Consortium for the
Participating Institutions. The Participating Institutions are the
American Museum of Natural History, Astrophysical Institute Potsdam,
University of Basel, University of Cambridge, Case Western Reserve
University, University of Chicago, Drexel University, Fermilab, the
Institute for Advanced Study, the Japan Participation Group, Johns
Hopkins University, the Joint Institute for Nuclear Astrophysics, the
Kavli Institute for Particle Astrophysics and Cosmology, the Korean
Scientist Group, the Chinese Academy of Sciences (LAMOST), Los Alamos
National Laboratory, the Max-Planck-Institute for Astronomy (MPIA),
the Max-Planck-Institute for Astrophysics (MPA), New Mexico State
University, Ohio State University, University of Pittsburgh,
University of Portsmouth, Princeton University, the United States
Naval Observatory, and the University of Washington.

Funding for SDSS-III has been provided by the Alfred P. Sloan
Foundation, the Participating Institutions, the National Science
Foundation, and the U.S. Department of Energy Office of Science. The
SDSS-III web site is http://www.sdss3.org/.

SDSS-III is managed by the Astrophysical Research Consortium for the
Participating Institutions of the SDSS-III Collaboration including the
University of Arizona, the Brazilian Participation Group, Brookhaven
National Laboratory, Carnegie Mellon University, University of
Florida, the French Participation Group, the German Participation
Group, Harvard University, the Instituto de Astrofisica de Canarias,
the Michigan State/Notre Dame/JINA Participation Group, Johns Hopkins
University, Lawrence Berkeley National Laboratory, Max Planck
Institute for Astrophysics, Max Planck Institute for Extraterrestrial
Physics, New Mexico State University, New York University, Ohio State
University, Pennsylvania State University, University of Portsmouth,
Princeton University, the Spanish Participation Group, University of
Tokyo, University of Utah, Vanderbilt University, University of
Virginia, University of Washington, and Yale University.

\facility{IRTF (SpeX)}

\software{emcee \citep{emcee}, Cloudy \citep{ferland13},
  pymccorrelation \citep{privon20}, mlinmix\_err
  \citep{kelly07}, Sherpa
  \citep{freeman01}, SimBAL \citep{leighly18}}, GMM  \citep{mg10}



\end{document}